\documentclass[final,1p,times]{elsarticle}

\usepackage{amssymb}
\usepackage{amsmath}
\usepackage{soul}
\usepackage{xcolor}
\usepackage[normalem]{ulem}

\journal{arXiv}

\begin{document}

\begin{frontmatter}

\title{Curvature-Guided Mechanics and Design of Spinodal and Shell-Based Architected Materials}

\author[inst1]{Somayajulu Dhulipala}

\affiliation[inst1]{organization={Department of Mechanical Engineering, Massachusetts Insitute of Technology},
            addressline={77 Massachusetts Ave}, 
            city={Cambridge},
            postcode={02139}, 
            state={Massachusetts},
            country={USA}}

\author[inst1]{Carlos M. Portela\corref{cor1}}
\ead{cportela@mit.edu}
\cortext[cor1]{Corresponding author}

\begin{abstract}
Additively manufactured (AM) architected materials have enabled unprecedented control over mechanical properties of engineered materials. While lattice architectures have played a key role in these advances, they suffer from stress concentrations at sharp joints and bending-dominated behavior at high relative densities, limiting their mechanical efficiency. Additionally, high-resolution AM techniques often result in low-throughput or costly fabrication, restricting manufacturing scalability of these materials. Aperiodic spinodal architected materials offer a promising alternative by leveraging low-curvature architectures that can be fabricated through techniques beyond AM. Enabled by phase separation processes, these architectures exhibit tunable mechanical properties and enhanced defect tolerance by tailoring their curvature distributions. However, the relation between curvature and their anisotropic mechanical behavior remains poorly understood.
In this work, we develop a theoretical framework to quantify the role of curvature in governing the anisotropic stiffness and strength of shell-based spinodal architected materials. We introduce geometric metrics that predict the distribution of stretching and bending energies under different loading conditions, bridging the gap between curvature in doubly curved shell-based morphologies and their mechanical anisotropy. We validate our framework through finite element simulations and microscale experiments, 
demonstrating its utility in designing mechanically robust spinodal architectures. This study provides fundamental insights into curvature-driven mechanics, guiding the optimization of next-generation architected materials for engineering applications. 
\end{abstract}


\end{frontmatter}

\section{Introduction}
The rapid advancement of additive manufacturing (AM) has revolutionized the development of architected materials, enabling the design of materials with unprecedented mechanical, electromagnetic, and optical properties \cite{kadic_3d_2019,surjadiPerspective2025}. These engineered materials, characterized by intricate micro- and nanoscale architectures, provide unique opportunities for tailoring mechanical behavior beyond the capabilities of conventional materials \cite{zheng_ultralight_2014, meza_strong_2014, bauer_approaching_2016, portela_extreme_2020, bauer_nanolattices_2017}. In the field of architected materials, research has primarily focused on optimizing mechanical properties relative to density, with a key metric being the scaling of stiffness and strength with relative density \cite{bauer_approaching_2016,crook_plate-nanolattices_2020,berger_mechanical_2017}. The scaling of stiffness $E^*$, approximated by $E^* \propto \overline{\rho}^{\alpha}$, where $\alpha$ is the stiffness scaling exponent has been attributed to the underlying microstructure across some relative density ranges \cite{gibson_cellular_1997}. Similarly, the scaling of strength, $\sigma^*_y$, has been captured by the form $\sigma^*_y \propto \overline{\rho}^{\beta}$, where $\beta$ is the strength scaling exponent. Early progress in the field of cellular materials extensively explored open-cell foams and linked their mechanical responses to scaling exponents of $\alpha = 2$ and $\beta = 1.5$ due to their bending-dominated deformation. However, advances in AM in the last 15 years have facilitated the exploration of new topologies that achieve improved scaling law exponents (closer to the theoretical limit of 1), steering research from readily available foams to periodic truss-based structures, and more recently to plate- and shell-based architected materials. Over the last few years, increased understanding of the influence of defects \cite{GLAESENER2023118918,LIU2017160, Defects2} in periodic architectures---inevitable in all AM processes---and the inherent issue of crack initiation and propagation via sharp junctions or nodes, has increased research interest in aperiodic and asymmetric architectures.

Truss-based architected materials have been extensively studied due to their ability to (via some morphologies) achieve stretching-dominated behavior, which enhances their mechanical efficiency and stiffness/strength scaling \cite{pellegrino_matrix_1986,DESHPANDE20011747}. The search for these ideal architectures has relied on defining a truss network in three dimensions that is statically and kinematically rigid by satisfying the condition  
\begin{equation}
    b - 3J+6 \geq 0,
\end{equation}
where $b$ represents the number of bars and $J$ denotes the number of non-foundational joints \cite{deshpande_foam_2001}. 
However, at manufacturable density regimes, these truss-based structures experience complex stress states at the nodes leading to stress concentration, poor scaling of strength and stiffness by deviating from the ideal slender-beam responses, and potential failure \cite{portela_impact_2018, hsieh_mechanical_2019,DefectsandStressConcentrations1,MEZA2017424}.

To overcome these limitations, researchers have explored plate-lattice-based architected materials, which enhance stretching-dominated behavior through closed-cell architectures. These structures exhibit near-optimal stiffness-to-weight scaling \cite{crook_plate-nanolattices_2020, berger_mechanical_2017, wang_achieving_2022} and leverage kinematic rigidity principles similar to truss-based lattices. However, stress concentrations due to abrupt cross-sectional changes still pose challenges, particularly for recoverability \cite{Dejean2018}. The introduction of Triply Periodic Minimal Surface (TPMS) structures—characterized by continuous, smooth curvature—has helped mitigate these issues. The near-zero mean curvature and negative Gaussian curvature of TPMS geometries promote a stretching-dominated response, leading to favorable stiffness and strength scaling \cite{bonatti_smooth-shell_2019,Shellular_2015,NGUYEN2017162,NickFangShell,ZHANG2018505}. Despite these advantages, TPMS-based architected materials rely on periodicity, making them highly sensitive to manufacturing defects, which can significantly degrade their mechanical performance \cite{LIU2017160, hsieh_mechanical_2019,DefectsandStressConcentrations1,Defects2, pham_damage-tolerant_2019}.

Beyond the inherent limitations of periodic architected materials, most existing architectures depend heavily on AM, which remains time-consuming and challenging for large-scale production. To address these drawbacks, aperiodic spinodal architected materials have emerged as a promising alternative. These materials exhibit low-curvature, stretching-dominated behavior in their shell-based form, and can be fabricated using methods beyond AM \cite{portela_extreme_2020, hsieh_mechanical_2019}. Among various approaches such as metallic foam dealloying \cite{Weismuller_Science_2021,hodge_nanoAg_2014,hodge_nanoCu,hodge_nanoCuAuPd}, one promising route for generating spinodal structures is polymerization-induced phase separation, which produces bicontinuous architectures in polymeric emulsions \cite{tsujioka_well-controlled_2008, bijels_2010,ericduns_natmat_2024}. Coating these structures via conformal deposition and subsequently removing the polymer matrix then yields thin-shell spinodal architected materials with enhanced mechanical properties \cite{portela_extreme_2020}.

While significant progress has been made in synthesizing self-architected spinodal structures, there still remain challenges in fully controlling the resulting microstructures (e.g., inducing anisotropy), and consequently, their anisotropic mechanical behavior remains poorly understood. One promising route for studying these materials is through computational modeling, where spinodal morphologies can be generated by simulating the phase-separation process via the Cahn-Hilliard phase field equation \cite{hsieh_mechanical_2019, vidyasagar_microstructural_2018} or via modeling of Gaussian random fields \cite{kumar_inverse-designed_2020, SOYARSLAN2018326}. By selecting an appropriate free-energy profile (or a level set to the random field), these computational models yield bicontinuous structures with near-zero mean curvature and negative Gaussian curvature. Additionally, anisotropic spinodal morphologies can be designed by introducing directional energy penalties (or constraining the directionality of wavevectors in the Gaussian random field), which influence mechanical anisotropy \cite{portela_extreme_2020, hsieh_mechanical_2019,THAKOLKARAN2025102274}. 

Despite these advances, there remains a fundamental gap in understanding the mechanics of shell-based spinodal architected materials—particularly in how anisotropic curvature distributions influence their mechanical response. While prior studies have emphasized the role of curvature in mechanical behavior \cite{PressurizedElasticSheets, taffetani_limitations_2019, wu_printed_2021, ramm_shell_2004,CurvatureSid}, a systematic framework linking curvature to mechanical anisotropy is still lacking. Recent data-driven approaches have been employed to optimize these morphologies for specific applications \cite{kumar_inverse-designed_2020, THAKOLKARAN2025102274, zheng_data-driven_2021,glaucio_2022}, but further fundamental mechanics-based understanding of curvature-induced anisotropy is needed.

In this work, we address this gap by advancing the theoretical understanding of curvature-dependent mechanical properties in shell-based spinodal architected materials. Specifically, we establish a geometric metric that quantifies anisotropic strength and stiffness, providing a predictive framework for optimizing mechanical performance across different loading conditions.

The manuscript is structured as follows: Section~\ref{sec:fabrication} describes the methods used to generate and fabricate the 3D morphologies of shell-based spinodal architected materials. Section~\ref{sec:CurvatureTheory} introduces existing curvature characterization metrics and explores their relation to mechanical responses. In Section~\ref{sec:Spinodal Mechanics}, we present a theoretical framework that predicts the distribution of stretching and bending energies across arbitrary loading directions. By simplifying this framework, we derive geometric metrics that capture the anisotropic stiffness of shell-based architected materials. Section~\ref{sec:FEADirStr} extends this framework to predict strength by constructing a 3D yield surface and validating it against finite element simulations. Finally, Section~\ref{sec:exptvalid} provides experimental validation of our theoretical framework by comparing predicted strength and stiffness metrics to performances obtained via uniaxial compression experiments on microscale spinodal morphologies.


\section{Generation of Spinodal Architectures}
\label{sec:fabrication}
\begin{figure}
\includegraphics[width = \textwidth]{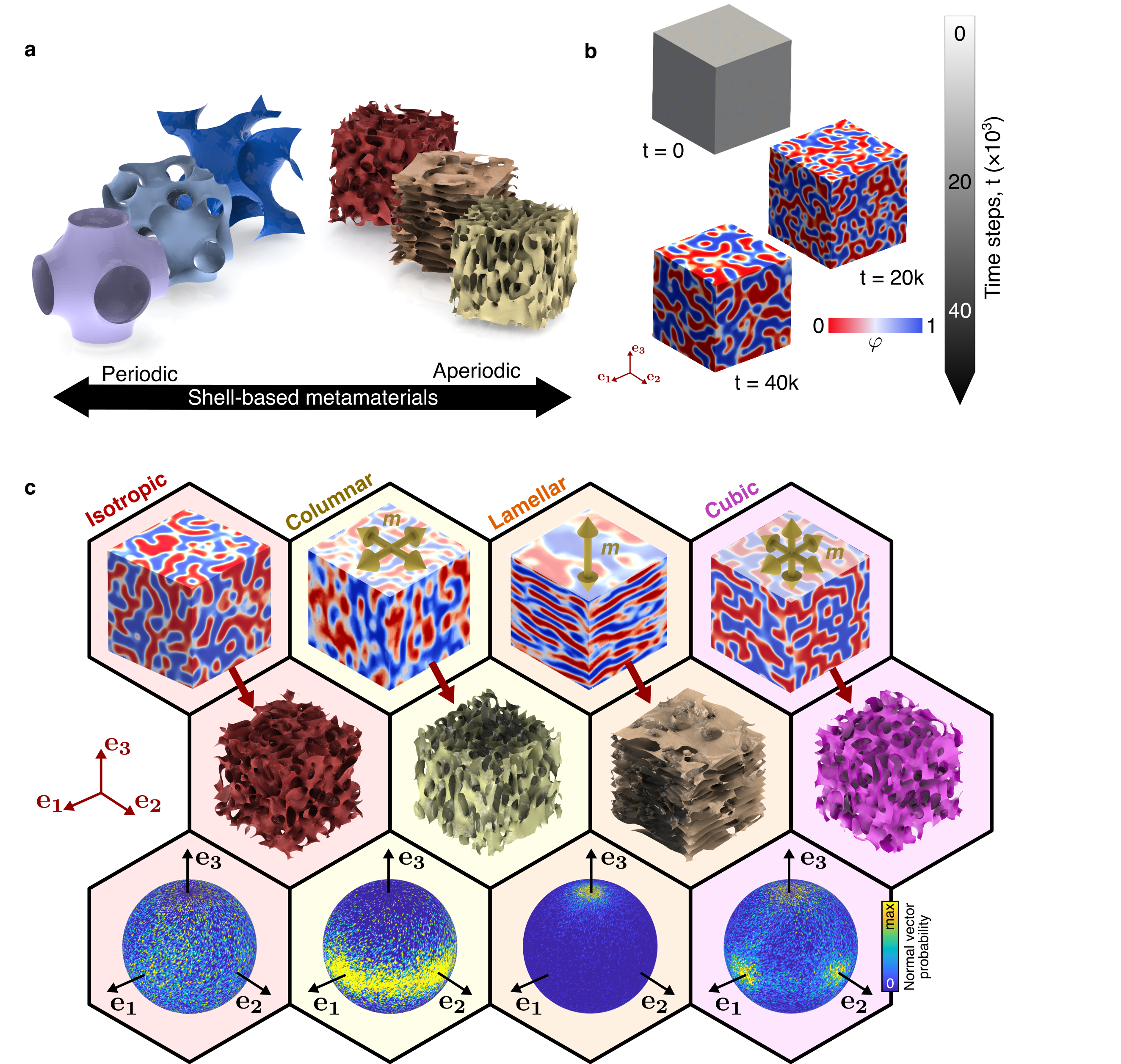}
\caption{(a) Renders of shell-based architected materials exhibiting a spectrum from periodic to aperiodic structures. (b) Temporal progression of the Cahn-Hilliard model depicting the phase separation in a two-phase mixture. (c) Spinodal morphologies—Isotropic, Columnar, Lamellar, and Cubic—produced by specifying preferential directions in the Cahn-Hilliard formulation, accompanied by pole figures illustrating the distributions of surface normals.}
\label{fig1:fab}
\end{figure} 

In this study, we primarily employ aperiodic spinodal morphologies, generated using a Cahn-Hilliard formulation\cite{vidyasagar_microstructural_2018}, as shown in Fig.~\ref{fig1:fab}a (right). To contrast these, we also consider triply periodic minimal surfaces (TPMS), referred to as periodic shell-based morphologies, generated using MS Lattice \cite{al-ketan_mslattice_2021} (Fig.~\ref{fig1:fab}a (left)). The aperiodic morphologies include isotropic, columnar, and lamellar variants (Fig.~\ref{fig1:fab}a (right)), whereas the periodic morphologies comprise three unique TPMS structures: p-cell, n14-1, and gyroid (Fig.~\ref{fig1:fab}a (left)).

Following the work by Vidyasagar et al., we computationally generate spinodal morphologies by formulating spinodal decomposition as an energy minimization problem where the chemical potential $\mu$ of a binary mixture is given by

\begin{equation}
\mu = \frac{1}{\epsilon} \left( \gamma(\mathbf{n})\frac{\partial B(\varphi)}{\partial\varphi} - \epsilon^2 (\gamma(\mathbf{n})\mathbf{\nabla}\varphi+|\nabla\varphi|(\mathbf{I}-\mathbf{n}\otimes\mathbf{n})\ \nabla_{\mathbf{n}}\ \gamma(\mathbf{n})) \right).
\end{equation}

Here, $\varphi$ represents the volume fraction of material ($\varphi=1$ for phase 1 and $\varphi=0$ for phase 2), and $\epsilon$ is the simulation length scale. The free energy for this binary mixture is given by the Ginzburg-Landau double-well potential

\begin{equation}
B(\varphi) = \frac{1}{4}\varphi^2(1-\varphi)^2.
\end{equation}

Anisotropy is introduced through the surface energy term $\gamma$

\begin{equation}
\gamma(\mathbf{n}) = \gamma_0\left[1-\sum_{i=1}^{\alpha}{a_i(\mathbf{n} \cdot \mathbf{m}_i)^{w_i}H(\mathbf{n}\cdot\mathbf{m}_i)}\right],
\end{equation}

where $\mathbf{m}_i$ represents a set of preferential directions, and $\mathbf{n}$ is the unit normal to the interface given by

\begin{equation}
\mathbf{n}(\mathbf{X},t) = \frac{\nabla\varphi(\mathbf{X},t)}{|\nabla\varphi(\mathbf{X},t)|}.
\end{equation}

The diffusion equation for the Cahn-Hilliard formulation is

\begin{equation}
\eta\frac{\partial\varphi}{\partial t} = \nabla\cdot\left[\sqrt{B(\varphi)}\nabla\mu\right],
\end{equation}

subject to periodic boundary conditions

\begin{equation}
\varphi(\mathbf{X}^+, t) = \varphi(\mathbf{X}^-, t).
\end{equation}

For simulation initialization, Gaussian randomization is applied, and the porosity is set to 50\%. These equations are solved using an FFT-based scheme with a time-step of $10^{-7}$ s while iterating over the phase field. Fig.~\ref{fig1:fab}b illustrates the evolution of morphology over several time steps without preferential directions.

Fig.~\ref{fig1:fab}c (top) displays morphologies obtained under different conditions: (i) Isotropic (no preferential direction), (ii) Columnar ($\mathbf{m} = \{\pm\mathbf{e_1},\pm\mathbf{e_2}\}$), (iii) Lamellar ($\mathbf{m} = \{\pm\mathbf{e_3}\}$), and (iv) Cubic ($\mathbf{m} = \{\pm\mathbf{e_1},\pm\mathbf{e_2},\pm\mathbf{e_3}\}$). The interfacial penalty induces preferential surface normals, evident in Fig.~\ref{fig1:fab}c (top).

To extract the interface between the two phases as a 3D shell morphology (represented by triangular mesh elements), a marching cubes algorithm is employed (Fig.~\ref{fig1:fab}c, middle), which demonstrates the use of interfacial penalties to tune curvature distributions and thereby mechanical properties. 

The change in curvature is visualized through pole figures (Fig.~\ref{fig1:fab}c, bottom), representing the density of surface normals. The normal vector probability is generated by discretizing a sphere into patches defined by 360 longitudes spanning the $\mathbf{e_1}$-$\mathbf{e_3}$ plane and 180 latitudes from the south to the north pole. For each morphology, the number of element normals falling within a given patch---weighted by the corresponding element area and normalized by the total surface area---is computed. Finally, this area-weighted normal count for each patch is normalized by the patch area to obtain the normal vector probability density.

Plotting the normal vector probability (Fig.~\ref{fig1:fab}c, bottom) reveals that the highest probability aligns with the preferred directions in the interfacial penalty function. For instance, the isotropic spinodal morphology exhibits randomly oriented normals, whereas the lamellar structure shows a higher density of normals along the $\pm\mathbf{e_3}$ directions.


\section{Curvature Theory Applied to a Generic Shell Element}
\label{sec:CurvatureTheory}

\begin{figure}
\includegraphics[width = \textwidth]{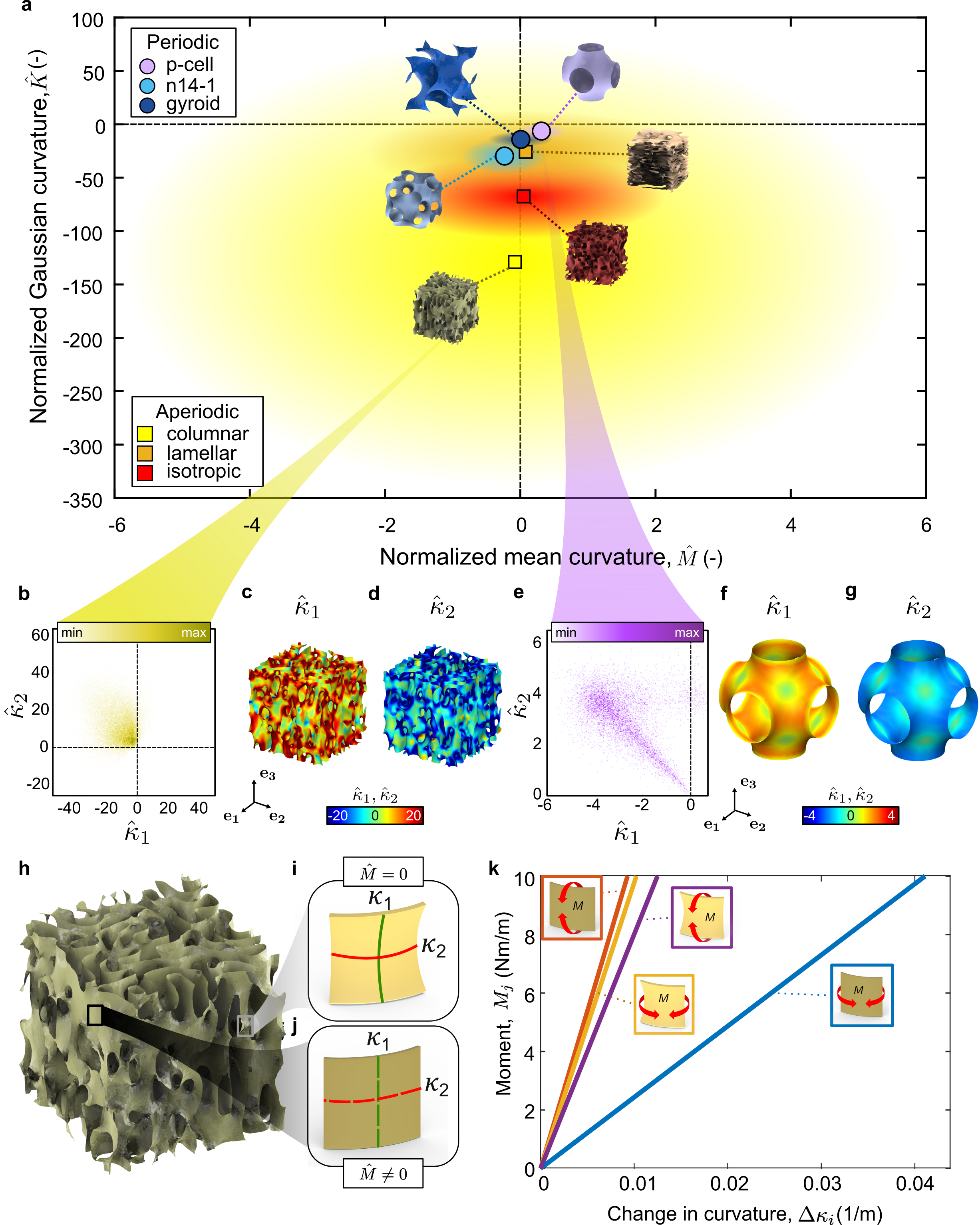}
\caption{(a) Normalized Gaussian curvature vs normalized mean curvature plotted for various shell-based architected materials. Shading around the data points represents the standard deviation.(b) Distribution of principal curvatures and 3D plots of (c) first and (d) second principal curvatures for a columnar spinodal morphology. (e) Distribution of principal curvatures and 3D plots of (f) first and (g) second principal curvatures for a p-cell morphology. (k) Moment vs change in curvature plot for shell sections with (i) zero mean curvature (\(\hat{M} = 0\)) and (j) non-zero mean curvature (\(\hat{M} \neq 0\)), extracted from (h) a columnar spinodal morphology, loaded laterally and longitudinally. }
\label{fig2:curv}
\end{figure}

The generated shell-based spinodal and TPMS morphologies are characterized by their zero-mean curvature and negative Gaussian curvature. This is illustrated in Fig.~\ref{fig2:curv}a, where the unit-cell size-normalized mean curvature \(\hat{M}\) and the Gaussian curvature \(\hat{K}\), are presented for three periodic morphologies (p-cell, n14-1, and gyroid) and three aperiodic morphologies (columnar spinodal, lamellar spinodal, and isotropic spinodal) of shell-based structures. The average mean and Gaussian curvatures of these morphologies lie on the \(\hat{M} = 0\) and \(\hat{K} < 0\) line, with the aperiodic spinodal architected materials exhibiting a larger spread in distribution.

Upon examining the principal curvature distribution plots of the columnar morphology (Fig.~\ref{fig2:curv}b) and a p-cell (Fig.~\ref{fig2:curv}e), we observe that the principal curvatures all lie in the second quadrant and on the line \(y=-x\). Furthermore, the 3D plots for the principal curvature (Figs.~\ref{fig2:curv}c, d, f, g) validate the presence of curvatures with differing signs. However, existing literature fails to capture an accurate interpretation of the mechanics of these shell-based morphologies beyond this point.

To further our understanding, we consider a columnar morphology (shown in Fig.~\ref{fig2:curv}h) with different curvatures displayed, revealing that all surface elements can be represented as shells. For our analysis, we consider two generic shell elements, one with $\hat{M}=0$ (Fig.~\ref{fig2:curv}i) and one with $\hat{M} \neq 0$ (Fig.~\ref{fig2:curv}j), corresponding to \emph{ideal} and \emph{non-ideal} design regions, respectively.
Using these generic \emph{ideal} and \emph{non-ideal} elements, we performed finite element analysis (FEA) to quantify the change in curvature of the shell sections upon application of a pure moment. Specifically, we explore two scenarios of moment application: lateral and longitudinal moment application, as shown in Fig.~\ref{fig2:curv}k.

These simple models highlight that for an ideal shell element ($\hat{M} = 0$), the bending stiffness in the principal direction---given by the slope of the moment about a perpendicular axis, $M_j$ vs. change in curvature $\Delta \kappa_i$ along the principal axis ($i,j = 1,2$)---is almost equivalent in both loading conditions (lateral ($i=2,j=1$) and longitudinal ($i=1,j=2$)). However, for the non-ideal element, we see that the bending stiffness for lateral loading is 4 times lower than the bending stiffness for longitudinal loading. This difference in stiffness for the non-ideal element can be understood by the change in Gaussian curvature induced by the moment \cite{taffetani_limitations_2019, wu_printed_2021,PressurizedElasticSheets, pini_how_2016,witten_stress_2007, CurvatureSid}. For the non-ideal element, a lateral moment causes $\kappa_2$ to change, but the overall Gaussian curvature of the element remains 0 despite this (since $\kappa_1=0$). However, a longitudinal moment causes $\kappa_1$ to change, and hence the overall Gaussian curvature changes as well. A change in Gaussian curvature in shells is attributed to a stretching response, which is energetically more costly than a bending response, leading to stiffening\cite{taffetani_limitations_2019, pini_how_2016}. This stiffening, represented as $\Delta S$, for a shallow shell with principal curvatures $\kappa_1$ and $\kappa_2$ is given by \cite{pini_how_2016}

\begin{equation}
\frac{\Delta S}{S_0} = \frac{b^4}{60h^2} (\kappa_y - \nu \kappa_x)^2 = f\left(\frac{W_s}{W_b}\right),     
\end{equation}

\noindent where $S_0$ is the stiffness of an uncurved plate, $b$ is the breadth of the plate, $h$ is the thickness of the plate, $\nu$ is the Poisson’s ratio, $\kappa_y$ is the lateral curvature, and $\kappa_x$ is the longitudinal curvature.

This increase is also dependent on the ratio of bending to stretching energy, $W_s/W_b$, confirming that whenever energy goes into a stretching mode, the structure becomes stiffer. Applying this to the shell elements considered here, we can integrate the membrane and bending energy densities over the surface of the shell for the different scenarios showcased. We find that the ratio of stretching to bending energy, $W_s/W_b$ for an ideal element loaded is 0.144 in both lateral and longitudinal loading scenarios, while a non-ideal element's energy ratio changes from 0.012 under lateral loading conditions to 0.690 in the longitudinal case. The unchanged energy ratio for the ideal element under both conditions indicates equivalent stretching-energy increases, hinting to higher mechanical efficiency under various loading scenarios. However, the non-ideal element's vastly different response for loading in different directions hints to an imparted directionality effect to a structure that is composed of several non-ideal elements.

In summary, while curvature can significantly stiffen shell surfaces, this stiffening effect is highly dependent on the orientation of the curvature with respect to the direction of loading. When a curved shell is loaded such that its Gaussian curvature changes, it leads to a stretching-dominated response and subsequent stiffening. In the next section, we apply this analysis to a generalized shell-based spinodal structure. This analysis, however, is not restricted solely to spinodal structures but rather can be applied to any shell-based morphology.


\section{Mechanics of Spinodal Morphologies - Stiffness}
\label{sec:Spinodal Mechanics}

The unique mechanical properties of shell-based spinodal architected materials are attributed to their stretching-dominated response, which is further linked to their curvature distribution \cite{portela_extreme_2020,guo_inverse_nodate}. Due to the diverse distribution of Gaussian curvature in differential shell elements within a spinodal structure (Fig.~\ref{fig2:curv}a),
any arbitrary loading inadvertently induces changes in Gaussian curvature in many of these shell elements, thus making the structures highly stretching-dominated. Moreover, the presence of certain preferential interface directions in anisotropic spinodal architectures results in specific loading scenarios causing larger changes in Gaussian curvature than others.

\begin{figure}
\includegraphics[width = \textwidth]{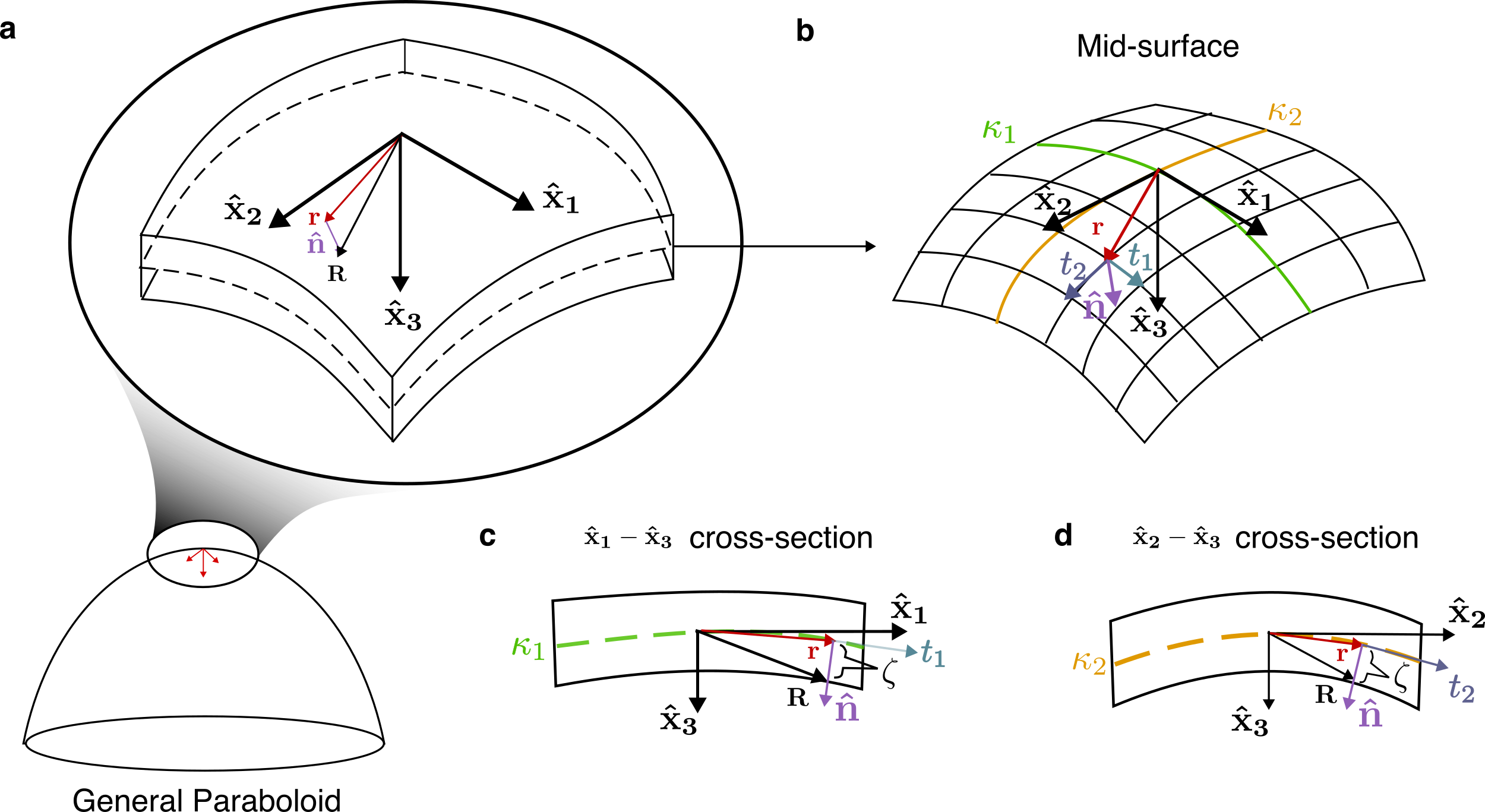}
\caption{(a) Schematic of a general paraboloid with an inset showing a magnified view near the origin. (b) Mid-surface representation of the shell geometry. (c) Cross-section of the shell in the $\hat{x}_1$–$\hat{x}_3$ plane. (d) Cross-section in the $\hat{x}_2$–$\hat{x}_3$ plane.}
\label{fig2a:kinematics}
\end{figure}

In this section, we propose a theoretical approach that effectively explains the stretching and bending behavior of shell-based architected materials based on their curvature distributions and loading directions. We consider a mesh of the spinodal structure and extract vertex normals and vertex curvatures and principal curvature directions \cite{rusinkiewicz_estimating_2004}. The midsurface of an arbitrary element on the mesh of a spinodal morphology can be approximated as a section of a paraboloid (Fig.~\ref{fig2a:kinematics}, \ref{sec:derivationparaboloid:appendix}) given by

\begin{equation}
\mathbf{r} = \left(\frac{\kappa_1{x_1}^2}{2} + \frac{\kappa_2{x_2}^2}{2}\right)\mathbf{\hat{x}_3} + x_1\mathbf{\hat{x}_1} + x_2\mathbf{\hat{x}_2}.
\end{equation}

Here, $\kappa_1$ and $\kappa_2$ represent the two principal curvatures, and $\mathbf{\hat{x}_1}$, $\mathbf{\hat{x}_2}$, and $\mathbf{\hat{x}_3}$ represent the principal curvature directions and the surface normal at $(x_1,x_2) = (0,0)$. The tangent vectors to the curve are given by

\begin{equation}
\mathbf{t_1} = \frac{\partial\mathbf{r}}{\partial x_1} = \kappa_1 x_1 \mathbf{\hat{x}_3} + \mathbf{\hat{x}_1}
\end{equation}

\noindent and

\begin{equation}
\mathbf{t_2} = \frac{\partial\mathbf{r}}{\partial x_2} = \kappa_2 x_2 \mathbf{\hat{x}_3} + \mathbf{\hat{x}_2},
\end{equation}

\noindent which reduce to $\mathbf{\hat{x}_1}$ and $\mathbf{\hat{x}_2}$ at the origin when $(x_1,x_2) \rightarrow (0,0)$. Correspondingly, the surface metric tensor $\mathbf{g}$ is defined as \cite{reddy_theory_2006_Ch11}

\begin{equation}
g_{ij} = \mathbf{t_i} \cdot \mathbf{t_j}.
\end{equation}

Thus, we can also define $a_{i} = \sqrt{g_{ii}} = \sqrt{\mathbf{t_{i}} \cdot \mathbf{t_{i}}}$, where $i = \{1,2\}$. For the case of $(x_1,x_2) \rightarrow (0,0)$, the surface metric tensor is a second-order identity matrix such that $a_1, a_2 = 1$. The normal vector to this surface is therefore given by

\begin{equation}
\mathbf{\hat{n}} = \frac{\mathbf{t_1} \times \mathbf{t_2}}{a_1 a_2},
\end{equation}
which also reduces to $\mathbf{\hat{x}_3}$ in the vicinity of the origin.

Considering the thickness of the shell, the coordinates of any arbitrary point on the shell are given by (Fig.~\ref{fig2a:kinematics}c,d)

\begin{equation}
\mathbf{R} = \mathbf{r} + \zeta \mathbf{\hat{n}},
\end{equation}
where $\zeta$ is the coordinate along $\mathbf{n}$. When $(x_1,x_2) \rightarrow (0.0)$, $\zeta$ becomes $x_3$, and $\mathbf{n}$ becomes $\mathbf{\hat{x}_3}$. 
Similar to the case of the surface, we can define a tangent to the thick shell close to the origin at any height $x_3$ as

\begin{equation}
\mathbf{T_i} = \frac{\partial\mathbf{R}}{\partial x_i} = \frac{\partial\mathbf{r}}{\partial x_i} + \zeta \frac{\partial\mathbf{\hat{n}}}{\partial x_i} + \mathbf{\hat{n}} \frac{\partial\zeta}{\partial x_i}, \quad \{i=1,2\}.
\end{equation}

This simplifies to $\mathbf{T_i} = \mathbf{t_i}(1 + x_3\kappa_i)$ for $i = \{1,2\}$. We can now define a global metric tensor, $\mathbf{G}$, given by

\begin{equation}
G_{ij} = \mathbf{T_i} \cdot \mathbf{T_j}.
\end{equation}

Subsequently, we define $A_1 = \sqrt{G_{11}} = \sqrt{\mathbf{T_1} \cdot \mathbf{T_1}} = a_1(1 + x_3\kappa_1)$, $A_2 = \sqrt{G_{22}} = \sqrt{\mathbf{T_2} \cdot \mathbf{T_2}} = a_2(1 + x_3\kappa_2)$, and $A_3 = 1$ \cite{reddy_theory_2006_Ch11}. In the vicinity of the origin, these reduce to $A_1 = 1 + x_3\kappa_1$, $A_2 = 1 + x_3\kappa_2$, and $A_3 = 1$ (i.e., $a_1=a_2=1$).

Next, we move on to setting up the mechanics problem. For a thin shell (following Kirchoff-Love theory), the displacements are of the form 

\begin{equation}
u_1(x_1,x_2,x_3) = u_{0,1}(x_1,x_2) + x_3\psi_1,
\label{eq:u1}
\end{equation}

\begin{equation}
u_2(x_1,x_2,x_3) = u_{0,2}(x_1,x_2) + x_3\psi_2,
\label{eq:u2}
\end{equation}

\begin{equation}
u_3(x_1,x_2,x_3) = u_{0,3}(x_1,x_2),
\label{eq:u3}
\end{equation}

\noindent where $u_1$, $u_2$, and $u_3$ are the displacements along the $x_1$, $x_2$, and $x_3$ directions; $u_{0,1}$, $u_{0,2}$, and $u_{0,3}$ are the mid-surface displacements that only depend on $x_1$ and $x_2$; and $\psi_1$ and $\psi_2$ are the rotations about the $\mathbf{\hat{x}_2}$ and $\mathbf{\hat{x}_1}$ axes, respectively. In a curvilinear coordinate system, the strains (under the small-strain approximation) are given by \cite{sokolnikoff_is_mathematical_1956,palazotto_nonlinear_1992}

\begin{equation}
    \varepsilon_{ii} = \frac{\partial}{\partial x_i}\left(\frac{u_i}{A_i}\right) + \frac{1}{A_i}\sum_{k=1}^{3}\frac{u_k}{A_k}\frac{\partial A_i}{\partial x_k},
    \label{eq:epsii}
\end{equation}

\begin{equation}
\varepsilon_{ij} = \frac{1}{2A_1A_2}\left[{A_i}^2\frac{\partial}{\partial x_j}\left(\frac{u_i}{A_i}\right) + {A_j}^2\frac{\partial}{\partial x_i}\left(\frac{u_j}{A_j}\right)\right]. 
\label{eq:epsij}
\end{equation}

\begin{figure}
\includegraphics[width = \textwidth]{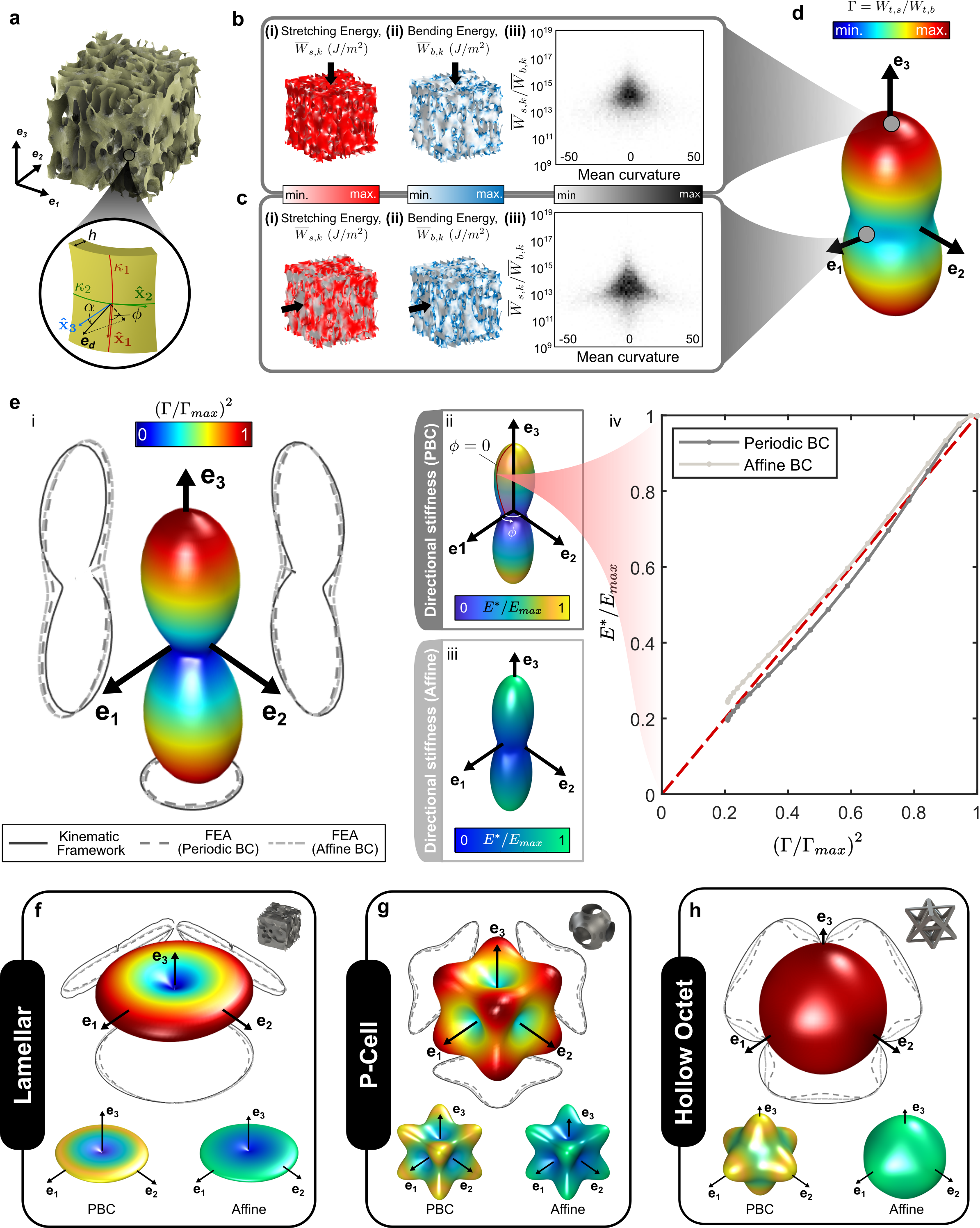}
\caption{(a) Schematic of mesh and important parameters of the mesh shown (inset). Theoretical (kinematic) prediction of the (i) stretching $(\overline{W}_{s,k})$ and (ii) bending energy $(\overline{W}_{b,k})$ and distribution of the ratio of stretching to bending energy $(\overline{W}_{s,k}/\overline{W}_{b,k})$ vs the mean curvature for loading along the unit vector (b) $[0,0,1]$, and (c) $[1,0,0]$. (d) 3D plot of total stretching to total bending energy $(\Gamma = \overline{W}_{t,s}/\overline{W}_{t,b})$ for a columnar geometry (e) (i) $\left(\Gamma/\Gamma_{max}\right)^2$  and normalized directional stiffness plots using (ii) periodic boundary conditions (PBC) and (iii) affine boundary conditions for columnar with (iv) normalized stiffness plotted against bending to stretching energy ratio for $\phi=0$ and $\theta=[0,90^\circ]$. Similar plots for (f) lamellar (g) p-cell and (h) hollow octet geometries.}
\label{fig3:Theo1}
\end{figure}

As shown in the schematic (Fig. \ref{fig3:Theo1}a), we specialize our analysis to affine deformation to the mid-surface in the direction $\mathbf{e_d}$ given by

\begin{equation}
    \mathbf{u_0} = U_g(\mathbf{r} \cdot \mathbf{e_d})\mathbf{e_d},
\end{equation}

\noindent where $U_g$ is the displacement gradient, and $\mathbf{r}$ is the position vector of any point on the mid-surface of the shell. The components of the displacement along the principal axes can be obtained by taking a dot product of $\mathbf{u_0}$ with the unit vectors along the principal axes (i.e., $u_{0,1} = \mathbf{u_0} \cdot \mathbf{\hat{x}_1}$, $u_{0,2} = \mathbf{u_0} \cdot \mathbf{\hat{x}_2}$, and $u_{0,3} = \mathbf{u_0} \cdot \mathbf{\hat{x}_3}$).

Under the thin-shell assumption ($\kappa_1 h,\kappa_2 h<0.1$), we can neglect the transverse shears ($\gamma_{i3}$), which allows us to approximate $\psi_1$ and $\psi_2$ as

\begin{equation}
    \psi_1 = -u_{0,1}\kappa_1 - \frac{\partial u_{3.0}}{\partial x_1},        
\end{equation}

\begin{equation}
    \psi_2 = -u_{0,2}\kappa_2 - \frac{\partial u_{0.3}}{\partial x_2}.        
\end{equation}

Thus, substituting the displacements (Eqs.~(\ref{eq:u1}),(\ref{eq:u2}),(\ref{eq:u3})) into the equations for strain (Eqs.~(\ref{eq:epsii}),(\ref{eq:epsij})), we obtain

\begin{equation}
    \varepsilon_{11} = \frac{U_g}{A_1}\left(\left(\mathbf{e_d} \cdot \mathbf{\hat{x}_1}\right)^2 - x_3\kappa_1\left(1 - \left(\mathbf{e_d} \cdot \mathbf{\hat{x}_2}\right)^2\right)\right),
    \label{eq:e11}
\end{equation}

\begin{equation}
    \varepsilon_{22} = \frac{U_g}{A_2}\left(\left(\mathbf{e_d} \cdot \mathbf{\hat{x}_2}\right)^2 - x_3\kappa_2\left(1 - \left(\mathbf{e_d} \cdot \mathbf{\hat{x}_1}\right)^2\right)\right),
    \label{eq:e22}
\end{equation}

\begin{equation}
    \varepsilon_{12} = \frac{U_g}{2}\left(\mathbf{e_d} \cdot \mathbf{\hat{x}_1}\right)\left(\mathbf{e_d} \cdot \mathbf{\hat{x}_2}\right)\left(\frac{1 - x_3\kappa_1}{A_2} + \frac{1 - x_3\kappa_2}{A_1}\right).
    \label{eq:e12}
\end{equation}

Assuming that the material is linear elastic with Young's modulus $E$ and Poisson's ratio $\nu$, the constitutive relation under a state of plane stress is given by

\begin{equation}
    \sigma_{11} = \frac{E}{1 - \nu^2}\left(\varepsilon_{11} + \nu \varepsilon_{22}\right),
    \label{eq:sig11}
\end{equation}

\begin{equation}
    \sigma_{22} = \frac{E}{1 - \nu^2}\left(\varepsilon_{22} + \nu \varepsilon_{11}\right),
    \label{eq:sig22}
\end{equation}

\begin{equation}
    \sigma_{12} = \frac{E}{1 + \nu}\varepsilon_{12}.
    \label{eq:sig12}
\end{equation}

The strain energy density per unit volume follows as $w=\frac{1}{2}\sum_{i,j}\sigma_{ij}\varepsilon_{ij}$, expressed as
\begin{equation}
    w = \frac{1}{2}\left(\sigma_{11}\varepsilon_{11} + \sigma_{22}\varepsilon_{22} + 2\sigma_{12}\varepsilon_{12}\right) = \frac{E}{2(1 - \nu^2)}\left[\varepsilon_{11}^2 + \varepsilon_{22}^2 + 2\varepsilon_{12}^2 + 2\nu\left(\varepsilon_{11}\varepsilon_{22} - \varepsilon_{12}^2\right)\right].
    \label{eq:W}
\end{equation}

Through this expression, the energy per unit area of the mesh can be obtained by integrating the energy through the thickness of the shells. This areal energy density, denoted by $\overline{W}$, is given by
\begin{equation}
    \overline{W} = \int_{-h/2}^{h/2} w A_1 A_2 \,dx_3,
\end{equation}

\noindent where $A_1$ and $A_2$ are multiplied to account for the curvature of the shell when integrating over the thickness. Substituting Eqs.~(\ref{eq:e11}), (\ref{eq:e12}), and (\ref{eq:e22}) into Eq.~(\ref{eq:W}), and taking a binomial expansion of all terms in $x_3$ up to quadratic order and ignoring terms in $x_3$ (since these will vanish when integrated from $-h/2$ to $h/2$), we obtain 
\begin{equation}
    \overline{W} = \overline{W}_s + \overline{W}_b,
\end{equation}

\noindent where $\overline{W_s}$ is the areal stretching energy density, given by 
\begin{equation}
    \overline{W}_s = U_g^2 \frac{Eh}{2(1-\nu^2)} \left(1 - (\mathbf{e_d} \cdot \mathbf{\hat{x}_3})^2\right)^2,
    \label{eq:Ws}
\end{equation}

\noindent and $\overline{W_b}$ is the areal bending energy density, given by
\begin{equation}
\begin{split}
    \overline{W}_b =  & U_g^2 \frac{Eh^3}{24(1-\nu^2)} \left[\left(2\kappa_1(\mathbf{e_d} \cdot \mathbf{\hat{x}_1})^2 + 2\kappa_2(\mathbf{e_d} \cdot \mathbf{\hat{x}_2})^2\right)^2\right. \\
    & + (\mathbf{e_d} \cdot \mathbf{\hat{x}_3})^2 \left(4\kappa_1^2(\mathbf{e_d} \cdot \mathbf{\hat{x}_1})^2 + 4\kappa_2^2(\mathbf{e_d} \cdot \mathbf{\hat{x}_2})^2 + (\kappa_1^2 + \kappa_2^2)(\mathbf{e_d} \cdot \mathbf{\hat{x}_3})^2\right) \\
    & \left. - \kappa_1 \kappa_2 \left(\left(1 - (\mathbf{e_d} \cdot \mathbf{\hat{x}_3})^2\right)\left(3 - 5(\mathbf{e_d} \cdot \mathbf{\hat{x}_3})^2\right) + 2\nu (\mathbf{e_d} \cdot \mathbf{\hat{x}_3})^2\right)\right].
    \label{eq:Wb}
\end{split}
\end{equation}

Correspondingly, the areal stretching energy density scales linearly with thickness $h$, while the areal bending energy density scales as $h^3$. For every $k$-th differential shell element under an arbitrary loading direction $\mathbf{e_d}$, we can calculate the product of the element area with the areal bending and stretching energy densities,  whose summation over all $N$ elements leads to the total bending and stretching energies,  $W_{t,b}$ and $W_{t,s}$, respectively, expressed as
\begin{equation}
    W_{t,b}=\sum_k^N \overline{W}_{b,k}S_k,
    \label{eq:Wbsum}
\end{equation}
\begin{equation}
    W_{t,s}=\sum_k^N \overline{W}_{s,k}S_k,
    \label{eq:Wtsum}
\end{equation}
where $\overline{W}_{s,k}$ and $\overline{W}_{b,k}$ are the stretching and bending energy per unit area, respectively, for the $k$-th element with corresponding area $S_k$. 

To validate the predictions of this kinematic framework, we applied it to a columnar spinodal morphology (Fig.~\ref{fig3:Theo1}a) with preferential orientation aligned with the $[0,0,1]$ direction, assuming linear-elastic constitutive properties (Young's modulus $E=2.7$ GPa, Poisson's ratio $\nu=0.2$). Upon uniaxial loading along the same $[0,0,1]$ direction, Fig.~\ref{fig3:Theo1}b presents the (i) bending  and (ii) stretching energies per unit area, clearly showing a stretching-dominated response. 
To ensure the architecture lies within the thin-shell regime (see \ref{sec:thickness:appendix}), we selected the uniform thickness of the shell to be $1/100 ^{\text{th}}$ of the unit cell size, ensuring that the product of the average maximum absolute curvature $\kappa \left( = \langle(\lvert\kappa_1\rvert+\lvert\kappa_2\rvert)/2\rangle\right)$ and the shell thickness $h$  falls between 0.1 and 0.01 ($0.01<\kappa h<0.1)$. The displacement gradient $U_g$, is taken to be 0.01. 

We observe that for loading in the $[0,0,1]$ direction, the stretching energy is uniform throughout the morphology, indicating efficient load transfer, while the bending energy is minimal throughout. Furthermore, the distribution of the stretching- to bending-energy ratio $\overline{W}_{s,k}/\overline{W}_{b,k}$ in all differential shell elements shows that elements with a mean curvature close to 0 exhibit a high value of $\overline{W}_{s,k}/\overline{W}_{b,k}$, with high-ratio occurrences becoming more rare for elements with non-zero mean curvature (Fig.~\ref{fig3:Theo1}b,iii).

When changing the loading direction to the $[1,0,0]$ direction, Fig.~\ref{fig3:Theo1}c shows noticeable differences in the distributions of (i) stretching energy, (ii) bending energy, and (iii) the ratio of $\overline{W}_{s,k}/\overline{W}_{b,k}$. In contrast to the case of loading along $[0,0,1]$, we do not observe a uniform distribution of stretching energy throughout and also identify an absolute increase in bending energy. This is validated by Fig.~\ref{fig3:Theo1}c,iii, where the average magnitude of the $\overline{W}_{s,k}/\overline{W}_{b,k}$ ratio decreased by 66\% and the distribution also qualitatively changed. It is interesting to note that the values of $\overline{W}_{s,k}/\overline{W}_{b,k}$ remain high around the mean curvature of 0, indicating that the ideal elements---as discussed in the previous section---remain agnostic to the loading direction, while the non-ideal elements experience a drastic change in their energy distribution. Similar plots are shown along the most rigid and compliant directions for other shell-based architected materials in \ref{sec:usbyubvsmean:appendix}, showcasing the same trends.

Compiling the total stretching and bending energies over all possible loading directions---$\overline{W}_{t,s}$ and $\overline{W}_{t,b}$, respectively---enables the calculation of the direction-dependent energy ratio $\Gamma=\overline{W}_{t,s}/\overline{W}_{b,k}$ which can be respresented as a 3D surface (Fig.~\ref{fig3:Theo1}d). Upon comparison of $\left(\Gamma/\Gamma_{max}\right)^2$ with the elastic surface obtained from computational homogenization with both periodic and affine boundary conditions, a good match is observed in terms of the degree of anisotropy (Fig.~\ref{fig3:Theo1}e). By conducting this validation for various periodic and aperiodic shell-based architected materials, as depicted in Fig.~\ref{fig3:Theo1}e--h, we confirm $\Gamma$ to serve as a proxy for anisotropy within shell-based architected materials. 

Homogenization is performed using the finite element method (ABAQUS), applying both periodic boundary conditions (PBC) and the more restrictive affine boundary conditions onto a morphology represented by structural shell elements. The degree of anisotropy predicted by our framework is more conservative than these two standards but is still effective in capturing the anisotropic mechanics of shell-based architected materials. The framework successfully captures the anisotropy for lamellar and columnar morphologies, predicting that the $[0,0,1]$ direction is the stiffest in the case of columnar morphology, while the $[1,0,0]$ and $[0,0,1]$ directions are the stiffest in the case of the lamellar spinodal morphology (Fig.~\ref{fig3:Theo1}e). This analysis also extends to TPMS structures such as the p-cell (Fig.~\ref{fig3:Theo1}g), where the framework accurately predicts that the $\left[\frac{1}{\sqrt{3}},\frac{1}{\sqrt{3}},\frac{1}{\sqrt{3}}\right]$ loading direction has the highest stiffness. Furthermore, the framework also applies to hollow truss-based lattices, which are not shell-based architected materials in the strictest sense (i.e. $M \neq 0$). However, the fundamental mechanical element is still a shell, and thus the kinematic framework can predict the anisotropic responses of such architectures (Fig.~\ref{fig3:Theo1}h). In the case of a hollow octet, the kinematic framework predicts slight anisotropy, with the highest stiffness being along the $[1,0,0],[0,1,0],[0,0,1]$ directions. Although PBC homogenization predicts a distinct anisotropic response, affine boundary conditions also predict a near isotropic response, deeming the predictions of the theoretical framework accurate and dominated by the affine-deformation assumption.

From a computational perspective, our method shares the same algorithmic complexity as reduced-order integration-based homogenization. This equivalence was confirmed through simulations conducted on meshes with varying numbers of elements (see \ref{sec:convergence:appendix}). The computational time for both reduced-order integration-based finite element analysis (FEA) and our kinematic framework exhibits a linear dependency on the number of elements. Once computed, the geometric parameters derived from our kinematic framework can be leveraged to predict behavior for different modulus and Poisson ratios. In contrast, FEA requires a new simulation setup, particularly when varying Poisson's ratio. Additionally, our kinematic framework demonstrates much faster convergence, allowing it to work effectively with coarser meshes and thereby further increasing computational speed. These advantages make our kinematic framework a quick and efficient tool for predicting anisotropy in shell-based architected materials.
In the next section, we explore how we can obtain geometric insights and use them to analyze shell-based architected materials based solely on their geometric parameters. Additionally, we generate proxies for stretching and bending energies, which can serve as metrics to validate the anisotropic designs of these materials.

\subsection{Geometric Metrics for Shell-Based Architected Materials}
\label{sec:closerlook}
\begin{figure}
\includegraphics[width = \textwidth]{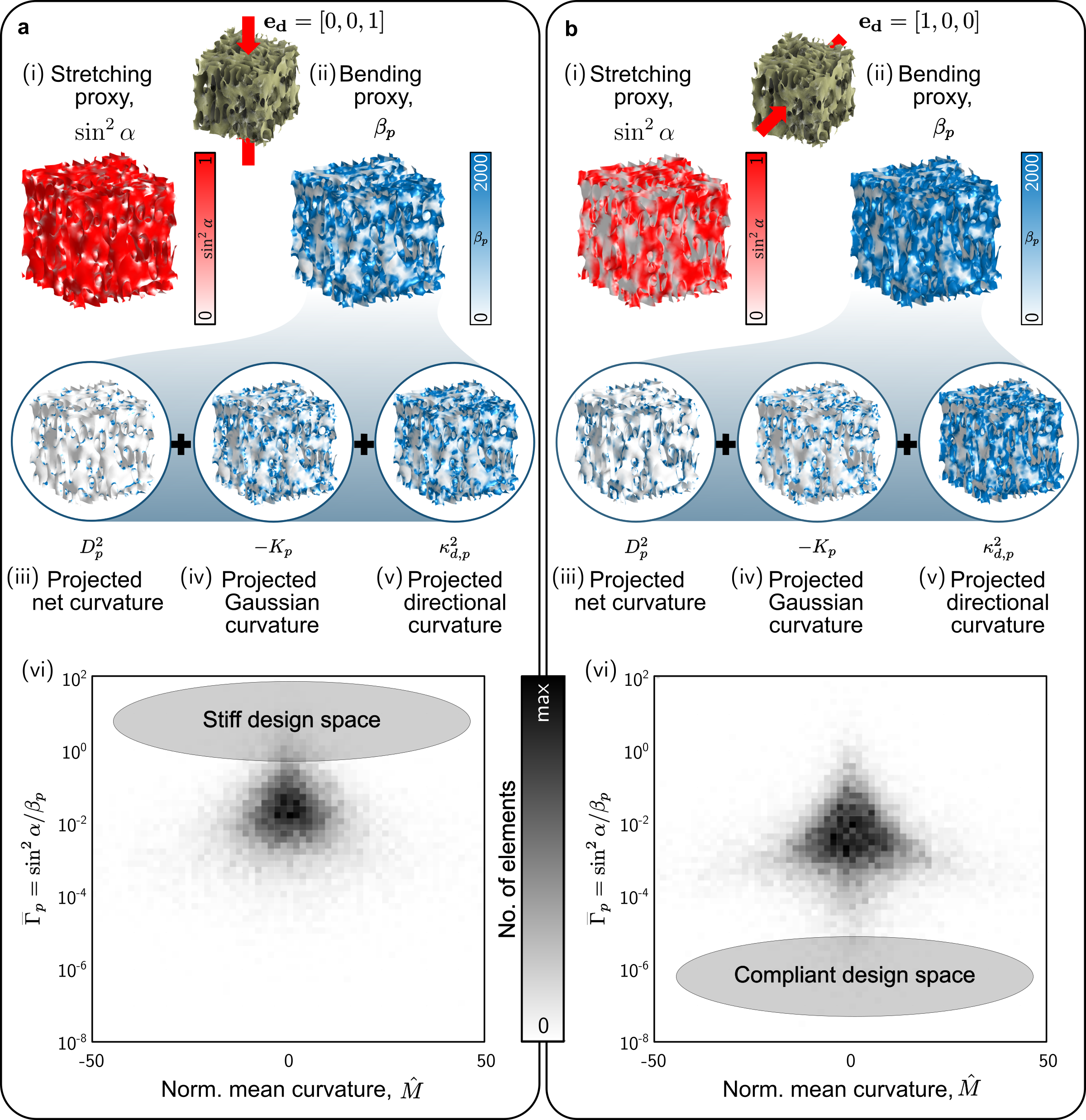}
\caption{(i) Stretching proxy, (ii) bending proxy, decomposed into (iii) projected net curvature, (iv) projected Gaussian curvature, and (v) projected directional curvature. (vi) A 2D histogram illustrating the local geometric proxy, $\overline{\Gamma_p} = \sin^2{\alpha}/\beta_p$, plotted against the normalized mean curvature for loading in the (a) $\mathbf{e_3}$ and (b) $\mathbf{e_1}$ directions.}
\label{fig4:Geom}
\end{figure}

In this section, we aim to develop a simple geometric metric that can be used to capture the anisotropic response of shell-based architected materials. Taking a closer look at Eqs.~(\ref{eq:Ws}) and ~(\ref{eq:Wb}), we observe that the displacement and material properties can be isolated from the geometric parameters. This isolation leads to the areal stretching energy density given by

\begin{equation}
    \overline{W}_s = \hat{f}\left(1 - (\mathbf{e_d} \cdot \mathbf{\hat{x}_3})^2\right) = \hat{f}\left(\sin^2{\alpha}\right),
\end{equation}

\noindent where $\alpha$, as depicted in Fig.~\ref{fig3:Theo1}a (inset), represents the angle between the loading direction and the surface normal. Fig.~\ref{fig4:Geom}i shows a plot of $\sin^2{\alpha}$ for a columnar spinodal morphology for loading in (a) $[0,0,1]$ and (b) $[1,0,0]$ directions.

Similarly, the areal bending energy density can be expressed as
\begin{equation}
\begin{split}
    \overline{W}_b & = \hat{g}\left(\kappa_d, \kappa_{d^2}, D, K, \alpha, \nu\right) \\
    &= \hat{g}\left(4\sin^2{\alpha}\left[\kappa^2_d\sin^2{\alpha}+\kappa_{d^2}\cos^2{\alpha}\right] \right. \\
    & \left. + 2D^2\cos^4{\alpha} - K\left[\sin^2{\alpha}\left(3-5\cos^2{\alpha}\right) + 2\nu\cos^2{\alpha}\right]\right) \\
    &= \hat{g}\left(\kappa^2_{d,p} + D^2_p - K_p\right),
    \label{eq:WBg}
\end{split}
\end{equation}
where $\kappa_d = \kappa_1\cos^2{\phi}+\kappa_2\sin^2{\phi}$ is the directional curvature given by Euler's formula, $\kappa_{d^2} = \kappa^2_1\cos^2{\phi}+\kappa^2_2\sin^2{\phi}$, $D=\sqrt{\frac{\kappa^2_1+\kappa^2_2}{2}}$ is the net curvature, and $K=\kappa_1\kappa_2$ is the Gaussian curvature. Here, $\phi$ is the angle between the projection of $\mathbf{e_d}$ in the $\mathbf{\hat{x}_1}-\mathbf{\hat{x}_2}$ plane and $\mathbf{\hat{x}_1}$.

Note from Eq.~(\ref{eq:WBg}) that the bending energy cannot be completely isolated from the material properties and is dependent on the Poisson's ratio, $\nu$, although in most practical cases (i.e., $\nu\sim 0-0.3$), this dependence is weak. 

From a geometric point of view, the terms in brackets in Eq.~(\ref{eq:WBg}) represent three important geometric features. The first term (shown in Fig.~\ref{fig4:Geom}v), 
\begin{equation}
    \kappa^2_{d,p}=4\sin^2{\alpha}\left[\kappa^2_d\sin^2{\alpha}+\kappa_{d^2}\cos^2{\alpha}\right],
\end{equation}represents a projected directional curvature metric. The second term (shown in Fig.~\ref{fig4:Geom}iii),\begin{equation}
    D^2_p=2D^2\cos^4{\alpha},
\end{equation}is the projection of the net curvature, and the third term (shown in Fig.~\ref{fig4:Geom}iv),\begin{equation}
    K_p=K\left[\sin^2{\alpha}\left(3-5\cos^2{\alpha}\right)+2\nu\cos^2{\alpha}\right],
\end{equation}is the projection of the Gaussian curvature. We define $\beta_p = \kappa^2_{d,p} + D^2_p - K_p$ as a proxy for bending energy. When plotting $\beta_p$ for a columnar spinodal morphology as shown in Fig.~\ref{fig4:Geom}a,ii and Fig.~\ref{fig4:Geom}b,ii---for loading along (a) $\left[0,0,1\right]$ and (b) $\left[1,0,0\right]$---it can be seen that of the three terms, the projected directional curvature metric $\kappa^2_{d,p}$ has the highest contribution to the bending proxy while the contribution of the other two terms less dominant. In particular, the contribution of the projected net curvature $D^2_p$ is minimal in both cases. 

Taking a ratio of the two proxies provides $\overline{\Gamma_p}=\sin^2{\alpha}/\beta_p$ which can be considered as a geometric proxy for $\overline{W}_s/\overline{W}_b$. Plotting this local geometric proxy $\overline{\Gamma_p}$ for different elements as a 2D distribution (Fig.~\ref{fig4:Geom}vi), we note that these plots are equivalent to those obtained in Figs. \ref{fig3:Theo1}b,iii and \ref{fig3:Theo1}c,iii. This similarity in the distributions validates the use of the geometric proxy as an alternative to the stretching-to-bending energy ratio to efficiently characterize the anisotropic response of a given shell-based architected material. For a given material morphology in a determined loading direction, more points in a region of high $\sin^2{\alpha}/\beta_p$ indicated more efficient loading and thereby a stiffer response along that direction (Fig.~\ref{fig4:Geom}a,vi), while more points in a lower region of $\sin^2{\alpha}/\beta_p$ indicate a more compliant response (Fig.~\ref{fig4:Geom}b,vi). This geometric proxy therefore provides a simple mechanics-based design rule to optimize the surface features of shell-based architected materials, enabling control of elastic responses along specified directions. Furthermore, to capture the global response of a given metamaterial, as given by $\Gamma = W_{t,s}/W_{t,b}$, we can alternatively take the ratio of the weighted sum of  $\sin^2{\alpha}$ and $\beta_p$ in a particular loading direction
\begin{equation}
    \Gamma_p = \frac{\sum_k^N \sin^2{\alpha_k}S_k}{\sum_k^N \beta_{p,k}S_k},
\end{equation}
where $S_k$ is the area of each element and the subscript $k$ indicates parameters calculated for each $k$-th element for a loading direction $\mathbf{e_d}$. Thus, $\Gamma_p$ could be used to evaluate the stiffness of a particular shell-based morphology in several directions and obtain its degree of anisotropy relying solely on its mesh attributes without the need of solving a homogenization problem. Additionally, the geometric metric, $\Gamma_p$ can also be used in topology optimization as an objective function to maximize or minimize depending on the requirements for stiffness or compliance. 

In the next sections, we explore routes to extend this kinematic framework towards determining the anisotropic strength of shell-based architected materials. To validate our approach, we implement a method using finite element analysis (FEA) for assessing the directional strength of these materials.


\section{Mechanics of Spinodal Morphologies - Strength}
\subsection{FEA Directional Strength}
\label{sec:FEADirStr}
\begin{figure}
\includegraphics[width = \textwidth]{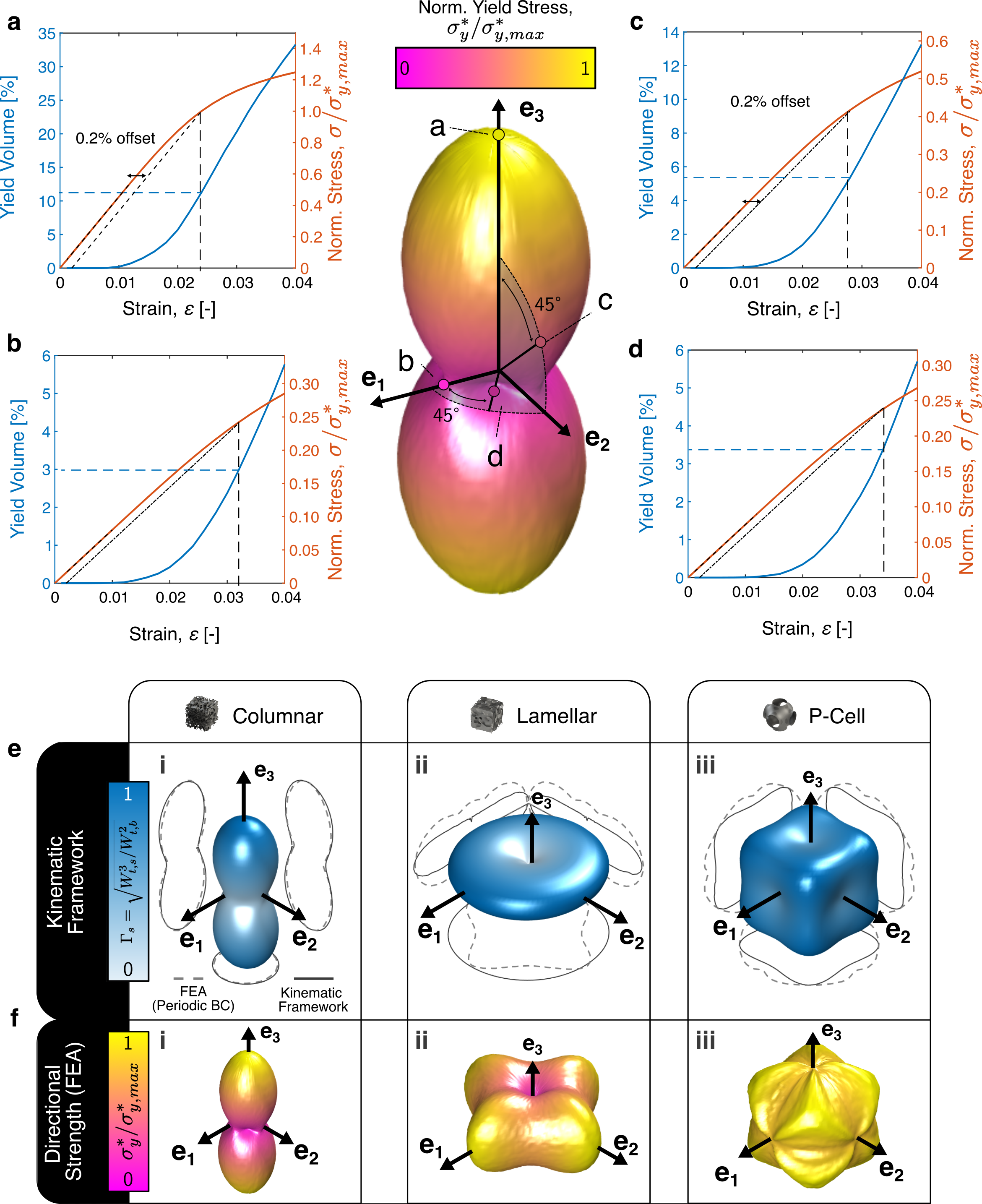}
\caption{Directional strength of a columnar spinodal morphology with stress-strain plots obtained from implicit FEA simulations for displacement in the (a) \(\hat{\mathbf{n}}=[0,0,1]\) direction, (b) [1 0 0] direction, (c) \(\left[0,\frac{1}{\sqrt{2}},\frac{1}{\sqrt{2}}\right]\) direction, and (d) \(\left[\frac{1}{\sqrt{2}}, \frac{1}{\sqrt{2}},0\right]\) direction. Prediction of effective directional strength from (e) the kinematic framework and (f) FEA for (i) columnar spinodal morphology, (ii) lamellar spinodal morphology, and (iii) p-cell.}
\label{fig5:dsirecstrength}
\end{figure}
Before attempting to theoretically predict the anisotropic strength of these materials, we first determine their anisotropic strength through the finite element method. Previous studies have examined the strength of spinodal architected materials \cite{hsieh_mechanical_2019} and the general anisotropic strength of shell-based architected materials using FEA \cite{bonatti_smooth-shell_2019}. Building on these efforts, we aim to extend the scheme used by Bonatti and Mohr \cite{bonatti_smooth-shell_2019} to predict the anisotropic strength of spinodal architected materials.

\sloppy To achieve this, we perform homogenization using periodic surface meshes of spinodal morphologies and triply periodic minimal surfaces with linear triangular shell elements (S3 elements, ABAQUS). We apply six linearly independent small-strain components, namely \(\varepsilon_{11}, \varepsilon_{22}, \varepsilon_{33}, \varepsilon_{12}, \varepsilon_{23},\varepsilon_{13}\) to a spinodal unit cell while imposing periodic boundary conditions. Through homogenization, we obtain both local stresses (at each element) and global stresses (for the entire unit cell), allowing us to calculate the local and global stiffness matrices via

\begin{equation}
    \mathbf{\Sigma} = \mathbb{C} \mathbf{\varepsilon},
\end{equation}
and
\begin{equation}
    \mathbf{\sigma}^{\left(\mathbf{i}\right)} = \mathbf{K}^{\left(\mathbf{i}\right)} \mathbf{\varepsilon},
\end{equation}

\noindent where \(\mathbf{\Sigma}\) is the global stress tensor, \(\mathbb{C}\) is the global stiffness tensor, \(\mathbf{\varepsilon}\) is the global strain tensor, \(\mathbf{\sigma}^{\left(\mathbf{i}\right)}\) is the local stress tensor at the \(i\)-th element, and \(\mathbf{K}^{\left(\mathbf{i}\right)}\) is the local stiffness tensor for the \(i\)-th element. To find the strength upon loading along a direction \(\mathbf{n}\), we apply a uniaxial compressive strain given by

\begin{equation}
    \left[\mathbf{\varepsilon_n}\right] = \left[\begin{matrix}
    \varepsilon_n & 0 & 0 \\
    0 & 0 & 0 \\
    0 & 0 & 0
    \end{matrix}\right],
\end{equation}

\noindent in an orthonormal basis where \(\mathbf{n}\) is one of the orthonormal basis directions. For this global strain, we can calculate the global stress and local stress as

\begin{equation}
    \mathbf{\Sigma} = \mathbb{C} \mathbf{Q} \mathbf{\varepsilon_n} \mathbf{Q}^T,
\end{equation}

\begin{equation}
    \mathbf{\sigma}^{\left(\mathbf{i}\right)} = \mathbf{K}^{\left(\mathbf{i}\right)} \mathbf{Q} \mathbf{\varepsilon_n} \mathbf{Q}^T.
\end{equation}

Here, \(\mathbf{Q}\) is the rotation tensor from the loading frame of reference to the unit cell frame of reference. We then vary the value of \(\varepsilon_n\) until 3\% of the volume of elements have reached the yield strength $\sigma_y$ of the constituent material. Given that we use shell elements, the von Mises stress is calculated under the assumption of plane stress. To find the value of $\varepsilon_n$ where 3\% of volume has yielded, we use a binary search algorithm since the yield percentage increases monotonically with \(\varepsilon_n\). We define the corresponding von Mises stress, \(\overline{\mathbf{\Sigma}}\), as our architected material's yield strength when loaded along a direction \(\hat{\mathbf{n}}\). Assuming a Young's modulus of 2.7 GPa, a Poisson's ratio of 0.2, and a yield strength of 68 MPa for the constituent material, we obtained the directional strength of a columnar morphology as shown in Fig.~\ref{fig5:dsirecstrength}. This 3D directional strength plot, obtained by sweeping through the solid angle comprising a sphere, depicts that the \(\mathbf{e_3}\) direction has the highest strength while the \(\mathbf{e_1}-\mathbf{e_2}\) plane has the lowest strength.

To validate the predictions from this approach, we also performed standard implicit simulations on unit cells of the columnar morphology with the same constituent material properties while using an elastic perfectly plastic material model. To probe directions that were off-axes from the principal directions, we extended our unit cells to a tessellation of \(3 \times 3 \times 3\) and rotated the tessellation such that the probing direction coincided with the \([0,0,1]\) direction. Following this rotation, we cut out a cube with side length equal to the unit cell size. Using free boundary conditions and imposing a displacement boundary condition on the top and a fixed boundary condition on the bottom, we simulated these meshes to capture the stress-strain behavior up to yield along four directions: (a) \(\hat{\mathbf{n}}=[0,0,1]\), (b) \(\hat{\mathbf{n}}=[1,0,0]\), (c) \(\hat{\mathbf{n}}=\left[0,\frac{1}{\sqrt{2}},\frac{1}{\sqrt{2}}\right]\), and (d) \(\hat{\mathbf{n}}=\left[\frac{1}{\sqrt{2}},\frac{1}{\sqrt{2}},0\right]\) (see Fig.~\ref{fig5:dsirecstrength}a--d)---producing a good match between the strength obtained through the aforementioned method and implicit simulations. Notably, the ratio of strength between the \([0,0,1]\) and \([1,0,0]\) directions was found to be $\sim$4 from the directional strength plot. Implicit simulations show that the strength obtained using a 0.2\% offset method in the \([1,0,0]\) direction (see Fig.~\ref{fig5:dsirecstrength}b) is one-fourth of the strength in the \([0,0,1]\) direction. However, we note that this method is limited due to its inability to capture buckling, especially at very low relative densities, thus restricting its application to the thin-shell regime (i.e., \(0.01 < \{\kappa_1 h, \kappa_2 h\} < 0.1\)).


\subsection{Theoretical Strength}
\label{sec:TheorStr}
To assess the strength of these morphologies, we consider the von Mises yield criterion. The distortion energy density for a $k$-th differential shell element in plane stress is given by

\begin{equation}
    w_{d,k} = \frac{1+\nu}{3E}\left[\sigma_{11}^2 - \sigma_{11}\sigma_{22} + \sigma_{22}^2 + 3\tau_{xy}^2\right].
\end{equation}

Substituting Eqs.~(\ref{eq:sig11}), (\ref{eq:sig22}), and (\ref{eq:sig12}), which provide all the stress components in the shell for a given displacement gradient $U_g$, provides an expression for the distortion energy $W_{d,k}$ of the shell element when integrated over its volume $V_k = S_k \int_{-h/2}^{h/2} A_1 A_2 dx_3$. Considering a first-order variation of the distortion energy (i.e., linearly varying with thickness), we obtain

\begin{equation}
    W_{d,k} = S_k \int_{-h/2}^{h/2} w_d A_1 A_2 dx_3 = U_g^2 S_k \frac{Eh(1+\nu^3)}{3(1-\nu^2)^2} \left(1 - (\mathbf{e_d} \cdot \mathbf{\hat{x}_3})^2\right)^2 + O[h^2].
\end{equation} 
The average von Mises stress in each element is then related to the distortion energy density as

\begin{equation}
    \langle \overline{\sigma}_k \rangle = \sqrt{\frac{3E w_{d,k}}{1+\nu}} = \frac{2 U_g E \left(1 - (\mathbf{e_d} \cdot \mathbf{\hat{x}_3})^2\right)}{1-\nu^2} \sqrt{\frac{(1+\nu^3)}{(12 + h^2 \kappa_1 \kappa_2)(1+\nu)}}.
\end{equation}

To find the yield strength in each direction $\mathbf{e_d}$, we increase $U_g$ until 3\% of the total volume has $\langle \overline{\sigma}_k \rangle \geq \sigma_y$, where $\sigma_y$ is the yield strength of the constituent material. In the limit where $h^2 \kappa_1 \kappa_2 \ll 1$, $\langle \overline{\sigma}_k \rangle$ simplifies to

\begin{equation}
    \langle \overline{\sigma}_k \rangle = \frac{2 U_g E \left(1 - (\mathbf{e_d} \cdot \mathbf{\hat{x}_3})^2\right)}{1-\nu^2} \sqrt{\frac{(1+\nu^3)}{12(1+\nu)}},
\end{equation}

\noindent indicating that $\langle \overline{\sigma}_k \rangle$ is highest when $\mathbf{e_d} \cdot \mathbf{\hat{x}_3} = 0$. This implies that if at least 3\% element volume is perpendicular to the loading direction (which is usually true in all loading directions) the limiting displacement  $U_{g,\text{lim}}$ is calculated simply by setting \(\mathbf{e_d} \cdot \mathbf{\hat{x}_3} = 0\), providing

\begin{equation}
    U_{g,\text{lim}} = \frac{\sigma_y (1-\nu^2)}{2E} \sqrt{\frac{12 (1+\nu)}{(1+\nu^3)}},
\end{equation}

\noindent which is only a function of material properties. Therefore, the total distortion energy at the macroscopic yield point can be estimated as

\begin{equation}
\begin{split}
    W_{t,d} &= \sum_k^N U_{g,\text{lim}}^2 S_k \frac{Eh(1+\nu^3)}{3(1-\nu^2)^2} \left(1 - (\mathbf{e_d} \cdot \mathbf{\hat{x}_3})^2\right)^2 \\
    &= \frac{\sigma_y^2 (1+\nu) h}{E} \sum_k^N S_k \left(1 - (\mathbf{e_d} \cdot \mathbf{\hat{x}_3})^2\right)^2.
\end{split}
\end{equation}

From the total distortion energy, the directional yield strength is

\begin{equation}
\begin{split}
    \sigma^*_y &= \sqrt{\frac{3E^* W_{t,d}}{UC^3 (1+\nu)}} \\
    &= \sqrt{\frac{3\sigma_y^2 E^* h}{UC^3 E} \left( \sum_k^N S_k \left(1 - (\mathbf{e_d} \cdot \mathbf{\hat{x}_3})^2\right)^2 \right)},
\end{split}
\end{equation}

\noindent where $E^*$ and $\sigma^*_y$ are the effective directional stiffness and strength, respectively. From Section~\ref{sec:Spinodal Mechanics}, we obtained that the directional stiffness is proportional to the square of the ratio of total stretching energy to total bending energy, $E^* \propto \Gamma^2 = \left(W_{t,s}/W_{t,b}\right)^2$. Additionally, from Eqs.~(\ref{eq:Ws}) and ~(\ref{eq:Wtsum}), we have

\begin{equation}
    \sum_k^N S_k \left(1 - (\mathbf{e_d} \cdot \mathbf{\hat{x}_3})^2\right)^2 = W_{t,s} \frac{2 (1-\nu^2)}{Eh U^2_{g,\text{lim}}},
\end{equation}

\noindent enabling us to further simplify,

\begin{equation}
    \sigma^*_y \propto \sqrt{\frac{6 \sigma_y^2 W_{t,s}^3 (1-\nu^2)}{W_{t,b}^2 UC^3 E^2 U^2_{g,\text{lim}}}} = \sqrt{\frac{2 (1+\nu^3) W_{t,s}^3}{UC^3 W_{t,b}^2 (1+\nu)(1-\nu^2)}} \propto \frac{W_{t,s}^{3/2}}{W_{t,b}}.
\end{equation}

We define this ratio as $\Gamma_s = \sqrt{W_{t,s}^3 / W_{t,b}^2}$, which acts as a proxy for directional strength. Figure \ref{fig5:dsirecstrength}e shows the theoretical directional strength for (i) columnar, (ii) lamellar, and (iii) p-cell morphologies, compared to FEA results in Fig. \ref{fig5:dsirecstrength}f. The trends are well captured by the theoretical model, though it fails to account for inhomogeneous deformations at larger strains. To incorporate these, we could vary the deformation gradient $U_g$ of each element by scaling with respect to its stiffness, though this would be computationally intensive. Excluding this, we focus on geometric aspects. A strength proxy $\Gamma_{s,p}$ is thus defined as

\begin{equation}
    \Gamma_{s,p} = \sqrt{\frac{W_{t,s}^3}{W_{t,b}^3}} = \frac{\left(\sum_k^N \sin^2{\alpha_k} S_k\right)^{3/2}}{\sum_k^N \beta_{p,k} S_k},
\end{equation}

\noindent where the quantities are as previously defined. This geometric proxy can be used for topology optimization from a strength perspective, where increasing $\Gamma_{s,p}$ in a particular direction leads to higher directional strength and vice versa.

\section{Experimental Validation}
\label{sec:exptvalid}
\begin{figure}[h!]
\includegraphics[width = \textwidth]{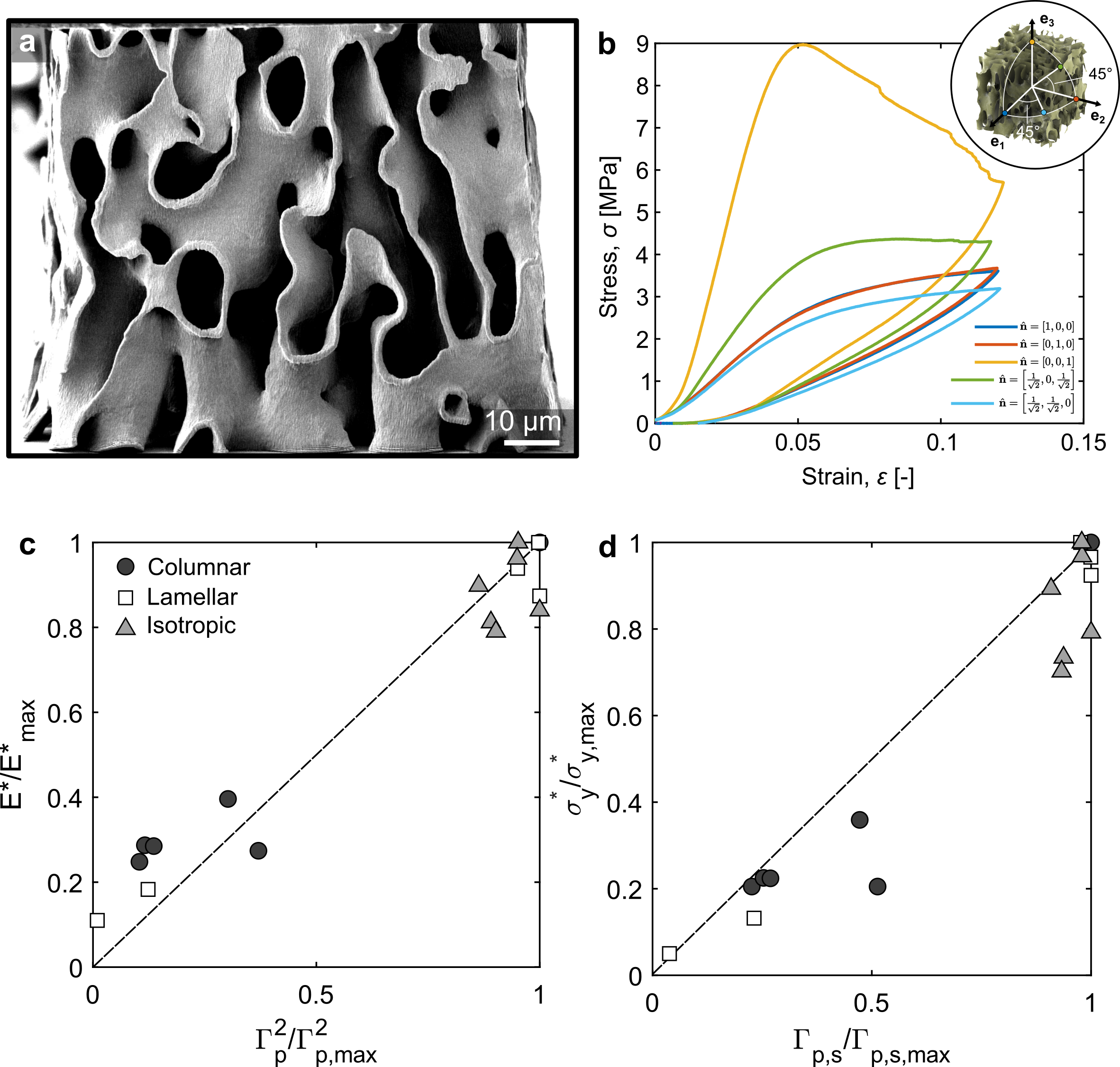}
\caption{(a) SEM image of a columnar morphology. (b) Stress-strain plots along different orientations for a columnar morphology with the probed directions indicated (inset). Scatter plots for columnar, lamellar, and isotropic spinodal morphologies of (c) normalized effective stiffness vs. normalized geometric proxy squared, and (d) normalized effective yield strength vs. normalized strength proxy. }
\label{fig6:exptvalidation}
\end{figure}

Finally, to verify the theoretical framework, we fabricated microscale thin-shell spinodal architected materials and performed uniaxial compression experiments up to 10\% strain for different morphologies. We probed several directions by generating unit cells with principal axes aligned with the direction to be probed. This was done by extending the Cahn-Hilliard phase field to a 3$\times$3$\times$3 cube and rotating the cube such that the direction to be probed faced the $\mathbf{e_3}$ axis. A 1$\times$1$\times$1 cube was then extracted from this larger cube, and the marching cubes algorithm was applied to obtain a 3D geometry. The resulting geometries were thickened so that the ratio of nominal thickness to unit cell size was 0.01. The geometry was scaled up to a total unit cell size of 100 \textmu{}m, sliced with a 0.3 \textmu{}m layer height and 0.2 \textmu{}m of hatching distance. Spinodal morphologies were fabricated out of IP-Dip resin using a two-photon polymerization system (PPGT2, Nanoscribe GmbH), using a writing speed of 10 mm/s and laser power of 25 mW and were developed for 3 hours in propylene glycol methyl ether acetate (PGMEA) followed by Iso-propyl alcohol for 30 minutes. Finally, the samples were dried using critical point drying (Tousimis Autosamdri-931) to prevent warping due to capillary effects during drying. 
As shown in Fig.~\ref{fig6:exptvalidation}a for a representative columnar morphology, the resulting structures had a unit cell size of 81$\pm$6 \textmu{}m and a thickness of 1.2$\pm$0.3 \textmu{}m. 

Following fabrication, the morphologies were uniaxially compressed using an Alemnis ASA nanoindenter with a 400-\textmu{}m-diameter diamond flat-punch tip at a strain rate of $10^{-3}$ s$^{-1}$, 
 applying a displacement of 10 \textmu{}m (corresponding to a strain of $\sim$12\%). Figure~\ref{fig6:exptvalidation}b shows the resulting stress-strain plots for a columnar morphology probed along different directions, where the effective stiffness was calculated as the slope of the linear loading regime, and the effective yield strength was measured using the 0.2\% offset method. As seen with FEA, the $\hat{\mathbf{n}}=[0,0,1]$ direction is the stiffest, while the plane perpendicular to $\hat{\mathbf{n}}=[0,0,1]$ is the most compliant. 

By plotting the measured effective stiffness for the different morphologies in the various probed directions against the square of the geometric proxy for a given morphology in the probed direction, we observe a strong correlation, as they align along the $y=x$ line (see Fig.~\ref{fig6:exptvalidation}c). Furthermore, plotting the normalized effective yield strength against the normalized strength proxy discussed in Section \ref{sec:TheorStr} reveals a similarly strong correlation (see Fig.~\ref{fig6:exptvalidation}d). These experiments strenghten the argument for these geometric proxies as effective tools in providing a quick and facile routes to capture the elastic and plastic anisotropy of shell-based architected materials.

\section{Summary}

Spinodal architected materials offer the advantage of being synthesized without the need for additive manufacturing, enabling scalable fabrication of architected materials and also presenting a unique route for ultralightweight properties when used as a template for shell-based form factors \cite{portela_extreme_2020}. These morphologies, characterized by their surfaces' low mean curvature and negative Gaussian curvature, result in highly stretching-dominated behavior. In this study, we present a theoretical model based on shell theory to predict the distribution of energy between stretching and bending modes under different uniaxial loading directions, extending beyond just spinodal architectures but also to other forms of shell-based architected materials. Our model accurately predicts directional stiffness with reduced computational cost compared to conventional finite element analysis by eliminating the need to generate a stiffness matrix.

Our findings indicate that elements with highly negative Gaussian curvature and zero mean curvature consistently exhibit strong stretching behavior regardless of the loading direction, while other elements show direction-dependent stretching-to-bending energy ratios. From the theory, we extracted important geometric parameters that provide insights into the distribution of stretching and bending energy. These geometric proxies can serve as design parameters for characterizing the anisotropic deformation behavior of shell-based architected materials and, most importantly, can be used as objective functions for computational design techniques when tuning the directional behavior of shell-based architected materials.

Notably, spinodal morphologies exhibit a strong dominance of stretching, especially in the thin-shell limit, allowing us to evaluate their yield strength primarily from a stretching perspective. This enables us to extract their distortion energy solely in terms of in-plane stresses. Additionally, we implemented a method to determine the directional strength of spinodal morphologies using finite element analysis, which validated a theoretical framework for understanding the directional strength of spinodal morphologies---providing a solid foundation for understanding the anisotropic yielding of these complex architected materials. Lastly, we are also able to validate our kinematic frameworks with experiments done on additively manufactured thin-shell spinodal morphologies for both stiffness and strength.

Through the metrics developed here, we hope to enable future understanding of shell-based architected material responses in dynamic and finite-strain conditions. Additionally, future research will focus on extending these models to predict fracture and to accommodate thicker shells where shear effects become significant.

\appendix

\section{Paraboloid as a representation for a generic shell surface}
\label{sec:derivationparaboloid:appendix}
Let any generic curved surface be represented in the orthonormal basis $(\mathbf{\hat{x}_1},\mathbf{\hat{x}_2},\mathbf{\hat{x}_3})$ by the function
\begin{equation}
    \mathbf{r}=x_1\mathbf{\hat{x}_1}+x_2\mathbf{\hat{x}_2}+f(x_1,x_2)\mathbf{\hat{x}_3}
\end{equation}
Performing a Taylor-series expansion of this function about $(x_1,x_2)=(0,0)$, we obtain
\begin{align}
    \mathbf{r} = &x_1\mathbf{\hat{x}_1}+x_2\mathbf{\hat{x}_2}+\\
    &\left(f(0,0)+f_{,x_1} (0,0)x_1+f_{,x_2} (0,0)x_2+f_{,x_1x_1}(0,0)\frac{x^2_1}{2}+f_{,x_2x_2}(0,0)\frac{x^2_2}{2}+f_{,x_1x_2}(0,0)x_1x_2+...\right)\mathbf{\hat{x}_3}.
\end{align}

If we set our origin such that $f(0,0)=0$ and also select the plane of our $\mathbf{\hat{x}_1},\mathbf{\hat{x}_2}$ basis such tangent to the curve (i.e., $f_{,x_1} (0,0)=0,f_{,x_2} (0,0)=0$), we can simplify this to
\begin{equation}
    \mathbf{r} = x_1\mathbf{\hat{x}_1}+x_2\mathbf{\hat{x}_2}+\left(f_{,x_1x_1}(0,0)\frac{x^2_1}{2}+f_{,x_2x_2}(0,0)\frac{x^2_2}{2}+f_{,x_1x_2}(0,0)x_1x_2+...\right)\mathbf{\hat{x}_3}.
\end{equation}
Next, if we rotate $\mathbf{\hat{x}_1}$ and $\mathbf{\hat{x}_2}$ in the same plane so that $f_{,x_1x_2}(0,0)=0$, this further simplifies to
\begin{equation}
    \mathbf{r} = x_1\mathbf{\hat{x}_1}+x_2\mathbf{\hat{x}_2}+\left(f_{,x_1x_1}(0,0)\frac{x^2_1}{2}+f_{,x_2x_2}(0,0)\frac{x^2_2}{2}+...\right)\mathbf{\hat{x}_3}.
\end{equation}
Such a rotation is possible and the directions of the orthonormal bases correspond to the eigenvectors of the matrix
\begin{equation}
\mathbf{K} = \begin{bmatrix}
f_{,x_1x_1}(0,0) & f_{,x_1x_2}(0,0) \\
f_{,x_1x_2}(0,0) & f_{,x_2x_2}(0,0)
\end{bmatrix},
\end{equation}

which correspond to the principal curvatures $\kappa_1,\kappa_2$.
\section{Applicable Thickness Regimes}
\label{sec:thickness:appendix}
\begin{figure}[h!]
\includegraphics[width = \textwidth]{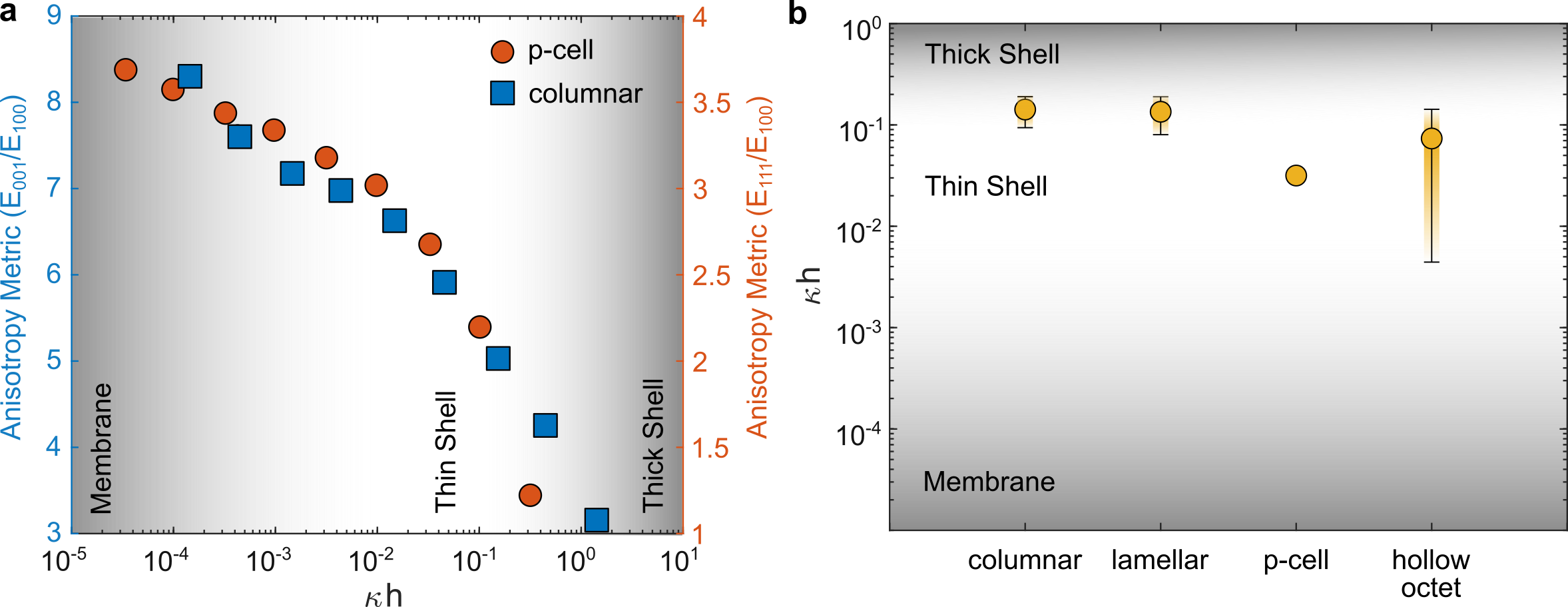}
\caption{(a) Variation of the maximum anisotropy metric of effective stiffnesses (i.e., the ratio of the highest to lowest effective stiffness) with the non-dimensionalized curvature \(\kappa h\). The predictions from the kinematic framework are most accurate in the thin shell regime ($0.01 < \kappa h < 0.1$). The shading indicates that the differentiation between regimes is blurred, as each morphology is composed of a range of curvatures. (b) The measured effective $\kappa h$ for the morphologies used in our study. The upper limit of the error bars corresponds to the effective mean of the maximum curvature of all elements, while the lower limit corresponds to the effective mean of the minimum curvature of all elements. All of our studied morphologies are approximately within the thin shell regime.}
\label{figb:thickness}
\end{figure}

Figure~\ref{figb:thickness}a shows the variation of FEA calculations (S3 elements, reduced-order integration with 3000 elements) of the ratio of maximum to minimum stiffness with a non-dimensional curvature \(\kappa h\) which is a product of the curvature \(\kappa\) and the shell thickness \(h\). The curvature \(\kappa\) used here is calculated by taking an area-weighted average of the mean absolute curvature of all the elements of the shell-based metamaterial $\left(\kappa = \langle(\lvert\kappa_1\rvert+\lvert\kappa_2\rvert)/2\rangle\right)$. For the columnar spinodal morphology, this corresponds to the ratio between the effective stiffness in the $[0,0,1]$ and the $[1,0,0]$ directions. For the p-cell morphology, this corresponds to the ratio between the effective stiffness in the $\left[\frac{1}{\sqrt{3}},\frac{1}{\sqrt{3}},\frac{1}{\sqrt{3}}\right]$ and the $\left[1,0,0\right]$ directions. 

The plots indicate that the anisotropic ratio for both geometries decreases with increasing thickness. As we progress from the thin-shell to thick-shell regime, a transition in scaling is observed for both geometries. Based on assumptions made when establishing the models, the kinematic framework should be applicable in the thin-shell regime (i.e., $0.01 < \kappa_1 h, \kappa_2 h < 0.1$). From Eqs.~(\ref{eq:Ws}) and ~(\ref{eq:Wb}), we see that the ratio $\Gamma = W_{t,sb}/W_{t,}$ varies inversely with $h^2$. However, when considering the anisotropy ratio between two loading directions, the $h$ terms cancel out, leaving a constant anisotropy ratio. Therefore, the theoretical framework presented in this work generates a single value of anisotropy irrespective of thickness, limiting its applicability to the thin-shell regime $0.01 < \kappa_1 h, \kappa_2 h < 0.1$.

The increase in anisotropy with decreasing thickness could potentially be explained by variations in the bending energy attributed to edge effects arising from the boundaries, which have a significant contribution in the membrane regime where the inherent bending energy may be negligible. At larger thicknesses, the role of shear cannot be neglected, contributing to the observed change in scaling. 

Figure~\ref{figb:thickness}b shows the non-dimensionalized curvature \(\kappa h\) for the morphologies used in this study. We see that the mean values are within the thin-shell regime. The maximum extents are calculated by taking an area-weighted mean of the absolute maximum curvature at each element, while the lower bounds are calculated by taking an area-weighted mean of the absolute minimum curvature at each element. The unshaded region represents the thin-shell regime. The boundaries between the regimes are intentionally blurred to capture the smooth transition.

\section{Stretch-to-Bend ratio for individual elements}
\label{sec:usbyubvsmean:appendix}

\begin{figure}[h!]
\includegraphics[width = \textwidth]{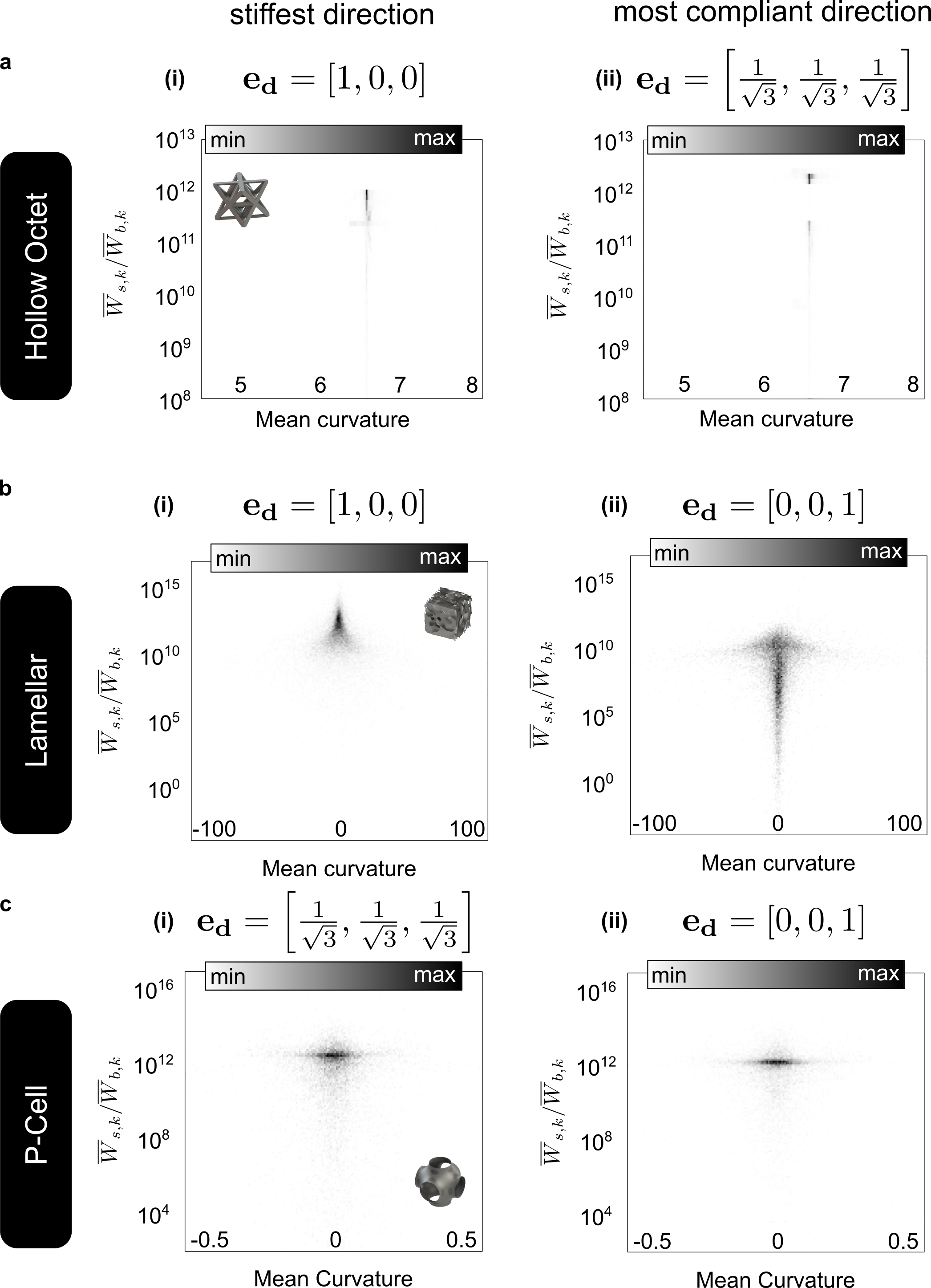}
\caption{Histogram of stretching-to-bending energies plotted for elements against the mean curvature in the (i) stiffest and (ii) most compliant directions for (a) Hollow Octet, (b) lamellar spinodal morphology, and (c) p-cell.}
\label{figb:usbyubvsmean}
\end{figure}

Stretch-to-Bend Ratios for individual elements are plotted in Fig.~\ref{figb:usbyubvsmean} for (a) hollow octet geometry, (b) lamellar spinodal morphology, and (c) p-cell in the stiffest and most compliant directions. For the hollow octet geometry, it is immediately noticeable that the mean curvature is not zero, highlighting the universality of the theoretical framework in providing predictions even for architected materials that are not strictly shell-based. 

In all geometries, there is a larger distribution of elements with higher values of $\overline{W}_{s,k}/\overline{W}_{b,k}$ along the stiffest direction. The subtle differences in distributions also indicate anisotropy. Additionally, there is a concentration of high $\overline{W}_{s,k}/\overline{W}_{b,k}$ values close to the zero mean curvature line, validating the idea that ideal elements (with zero mean curvature and negative Gaussian curvature) provide mechanical resilience to shell-based architected materials, while non-ideal elements induce anisotropy.

\section{Convergence and computational efficiency}
\label{sec:convergence:appendix}

Based on the order of operations, we calculated and compared computation times between FEA and the theoretical framework. The rate-limiting step for FEA is the inversion (and assembly) of the stiffness matrix, while for the theoretical framework, it is extracting the normals, principal curvatures and directions, and element areas from a mesh as vertex attributes. 

\begin{figure}[h!]
\includegraphics[width = \textwidth]{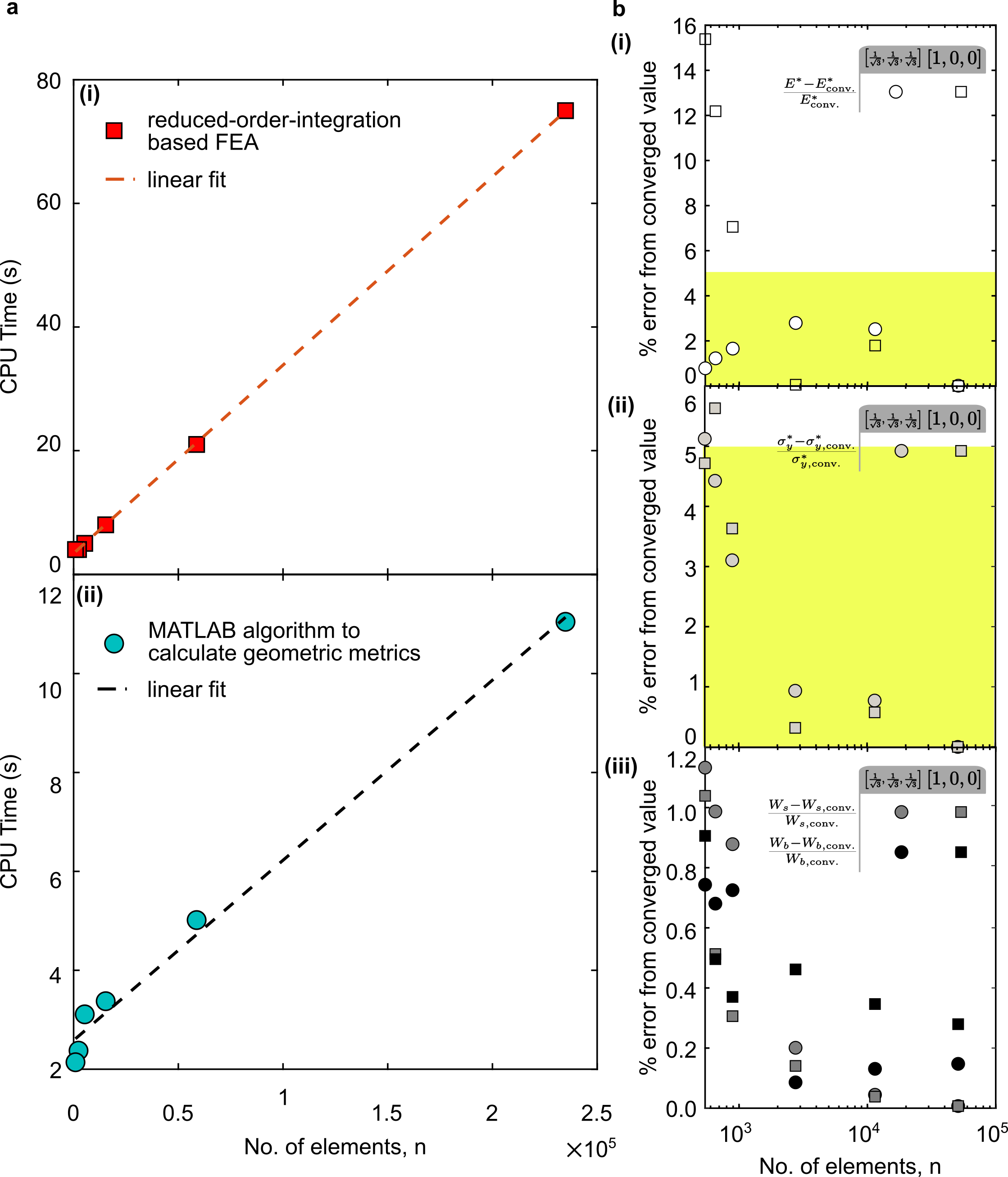}
\caption{ (a) Variation of CPU time with the number of elements for (i) reduced-order integration based FEA and (ii) extraction of geometric parameters from the mesh. 
    (b) Convergence studies performed on p-cell for (i) stiffness calculated from homogenization, (ii) strength calculated from homogenization, and (iii) stretching and bending energies calculated from geometric parameters extracted from the mesh.}
\label{figb:computation}
\end{figure}

Fig.~\ref{figb:computation}a,i shows the CPU time for running homogenization on surface meshes with different numbers of S3 shell elements using a reduced-order integration scheme. This also includes the time taken to match nodes to apply periodic boundary conditions. The CPU time increases linearly with the number of elements, indicating that FEA scales as $O(n)$. Additionally, Fig.~\ref{figb:computation}a,ii demonstrates that the computation time for geometric metrics also varies linearly with the number of elements, indicating that this also scales as $O(n)$. 

Performing convergence studies for the p-cell using periodic boundary conditions, we observed that the directional stiffness along $[1,0,0]$ and $\left[\frac{1}{\sqrt{3}},\frac{1}{\sqrt{3}},\frac{1}{\sqrt{3}}\right]$ directions converged ($<5\%$ error) at about 3000 elements (Fig.~\ref{figb:computation}b,i), particularly along the most compliant direction ($\mathbf{e_3}$). Similar convergence studies for the strength showed faster convergence at around 700 elements (Fig.~\ref{figb:computation}b,ii). Surprisingly, convergence for calculating stretching and bending energies extracted from the geometric metrics was reached quickly, showing a very low error from the very first mesh tested (Fig.~\ref{figb:computation}b,iii). 

These findings, coupled with the linear scaling of computation time with the number of elements, indicate that the theoretical framework can be a computationally efficient tool to analyze any shell-based structure for its anisotropy.

\section*{Acknowledgements}
Carlos M. Portela acknowledges financial support from the National Science Foundation (NSF), USA CAREER Award (CMMI-2142460), and the MIT MechE MathWorks Seed Fund and Engineering Fellowship Program, USA. This work was carried out in part through the use of MIT.nano’s facilities. 

  \newpage
 \bibliographystyle{elsarticle-num} 
 \bibliography{refsv2}

\end{document}